\documentclass[11pt,a4paper]{article}

\usepackage{booktabs}
\usepackage{natbib}
\usepackage{amsfonts}
\usepackage{amsmath}
\usepackage{graphicx}
\usepackage{algorithmic}
\usepackage{fancyvrb}
\usepackage[ruled,linesnumbered]{algorithm2e}

\SetAlFnt{\small}
\SetAlCapFnt{\small}
\SetAlCapNameFnt{\small}
\SetAlCapHSkip{0pt}
\DontPrintSemicolon
\usepackage{tikz,pgfplots}
\usepackage[hidelinks]{hyperref}
\usepackage[left=2.5cm,right=2.5cm,top=3cm,bottom=3cm]{geometry}
\usepackage{soul}
\usepackage[T1]{fontenc}

\newcommand{\Alin}{\mathcal{A}}
\newcommand{\Blin}{\mathcal{B}}
\newcommand{\Slin}{\mathcal{S}}

\renewcommand{\cite}[2][]{\citep[#1]{#2}}

\title{NEP: a module for the parallel solution of nonlinear eigenvalue problems in SLEPc\thanks{This work was supported by the Spanish Agencia Estatal de Investigaci\'{o}n under grants TIN2016-75985-P and PID2019-107379RB-I00 (including European Commission FEDER funds).}
}

\author{Carmen Campos\thanks{D.\ Sistemes Inform\`atics i Computaci\'o, Universitat Polit\`ecnica de Val\`encia, Val\`encia, Spain
  (\texttt{mccampos@dsic.upv.es}).}
\and Jose E. Roman\thanks{D.\ Sistemes Inform\`atics i Computaci\'o, Universitat Polit\`ecnica de Val\`encia, Val\`encia, Spain
  (\texttt{jroman@dsic.upv.es}).}
}

\begin{document}

\maketitle

\begin{abstract}
SLEPc is a parallel library for the solution of various types of large-scale eigenvalue problems. In the last years we have been developing a module within SLEPc, called NEP, that is intended for solving nonlinear eigenvalue problems. These problems can be defined by means of a matrix-valued function that depends nonlinearly on a single scalar parameter. We do not consider the particular case of polynomial eigenvalue problems (which are implemented in a different module in SLEPc) and focus here on rational eigenvalue problems and other general nonlinear eigenproblems involving square roots or any other nonlinear function. The paper discusses how the NEP module has been designed to fit the needs of applications and provides a description of the available solvers, including some implementation details such as parallelization. Several test problems coming from real applications are used to evaluate the performance and reliability of the solvers.
\end{abstract}

\section{Introduction}

SLEPc, the Scalable Library for Eigenvalue Problem Computations \citep{Hernandez:2005:SSF,Roman:2019:SUM2}, is a software package for the parallel solution of large-scale eigenvalue problems, with special focus on iterative methods to compute only part of the spectrum. It started as a collection of solvers for the linear eigenvalue problem, but its scope soon extended to other classes of related problems such as the singular value decomposition, the polynomial eigenvalue problem and the general nonlinear eigenvalue problem. SLEPc has a modular design, and this paper deals with the module called \texttt{NEP}, which is intended for the solution of general nonlinear eigenvalue problems. These eigenproblems depend nonlinearly on the eigenvalue parameter only, not the eigenvector. SLEPc also includes a basic nonlinear inverse iteration solver for problems that are nonlinear with respect to the eigenvector, but we do not discuss it here.

The \texttt{NEP} module can be used to solve rational eigenvalue problems, those defined by means of rational functions, or the more general class of nonlinear eigenproblems defined via arbitrary nonlinear functions (a formal definition will be given in \S\ref{sec:math}). Note that polynomial eigenvalue problems are excluded because they belong to PEP, a different SLEPc module, whose details were described for the most part in a previous paper \cite{Campos:2016:PKS}.

Nonlinear eigenvalue problems arise in many areas of computational science and engineering. For example, in wave phenomena eigenvalue problems become nonlinear when certain characteristic properties of the material depend on the eigenvalue in the frequency range of interest; a typical example is found in the analysis of photonic crystals, where the relative permittivity is modeled by a rational function of the eigenvalue (the complex-valued frequency) \cite{Brule:2016:CAC}. Another example appears in the spectral analysis of delay differential equations \cite{Michiels:2014:SCC}. In some applications, the nonlinear character comes from using special boundary conditions for the solution of differential equations \cite{Vandenberghe:2014:DBS,Beeumen:2018:CRM}.

The topic of nonlinear eigenvalue problems has been very active in the numerical linear algebra community in the last years. Many numerical methods have been proposed in the literature, which have been surveyed by \citet{Mehrmann:2004:NEP} and, more recently, by \citet{Guttel:2017:NEP}. The merit of some of these methods has been illustrated by solving various problems arising from real applications. However, as it often happens, these state-of-the-art algorithms are not available in the form of reliable and efficient free software implementations that can be widely used by the scientific computing community. To the best of our knowledge, apart from Matlab implementations of some algorithms, the only available package containing solvers for the general nonlinear eigenproblem is NEP-PACK \cite{Jarlebring:2018:NJP} for the Julia programming language. Our goal in SLEPc is to provide well-tested robust solvers with high performance for parallel computers that can be used from C/C++, Fortran and Python. We started the development of the \texttt{NEP} module in 2013, and since then it has been used in several scientific computing projects, e.g., in the simulation of photonic crystals \cite{Demesy:2020:NEP,Zolla:2018:PHD}, or the analysis of scattering resonances in metal-dielectric nano-structures \cite{Araujo:2020:CSR}.

In this paper we will describe the methods that are currently available in SLEPc's \texttt{NEP} module, and provide some relevant details of our implementation. But we will also discuss design decisions we have taken in order to organize the code. These decisions are usually geared towards making a friendly user interface. From the user point of view, defining a nonlinear eigenvalue problem is more difficult than in the linear case, since it is necessary to work with matrix-valued functions rather than with constant matrices. In SLEPc we provide ways to define these functions, and enable the solvers with capabilities to work with such functions, or their derivatives, and in some cases to compute matrix functions as well.

Rational eigenvalue problems are quite special because it is feasible to convert them to equivalent polynomial eigenvalue problems and it is also possible to solve them via linearization \cite{Su:2011:SRE}. But our approach is to use the same methods as for the more general nonlinear eigenproblems. 

The rest of the paper is organized as follows. Section~\ref{sec:math} gives a formal definition of the nonlinear eigenvalue problem and provides some necessary background. In section~\ref{sec:design} we will describe the design of the \texttt{NEP} module and the user interface. Section~\ref{sec:solvers} is devoted to the description of the different solvers available in SLEPc. A brief overview of the parallelization of the methods is included in section~\ref{sec:parallel}. The results of computational experiments are shown in section~\ref{sec:performance}. We wrap up with some concluding remarks in section~\ref{sec:concl}.

\section{Mathematical background\label{sec:math}}

Given a matrix-valued function $T:\Omega\rightarrow\mathbb{C}^{n\times n}$ defined in the open set $\Omega\subseteq\mathbb{C}$, the nonlinear eigenvalue problem (NEP) consists in determining eigenpairs $(x,\lambda)\in\mathbb{C}^n\times\mathbb{C}$ satisfying
\begin{equation}\label{eq:nep}
T(\lambda)x=0,\qquad x\neq0.
\end{equation}
The vector $x$ and the scalar $\lambda$ are referred to as eigenvector and eigenvalue, respectively. We will assume that $T$ is holomorphic in the domain $\Omega$ and that it is regular, that is, $\det T(z)$ is not identically zero. The eigenvalues are the solutions of the scalar equation $\det T(z)=0$. In addition to (right) eigenvectors, $x$, some applications also require the computation of left eigenvectors, $y$, satisfying
\begin{equation}\label{eq:nep-left}
y^*T(\lambda)=0^*,\qquad y\neq0,
\end{equation}
where $(\cdot)^*$ denotes the conjugate transpose of a vector or matrix. Left eigenvectors can be computed as right eigenvectors of $T^*$.

The function $T$ can always be expressed in \emph{split form}
\begin{equation}\label{eq:split}
T(z)=\sum_{i=1}^{\ell}A_if_i(z),
\end{equation}
for constant coefficient matrices $A_1,\dots,A_\ell\in\mathbb{C}^{n\times n}$ and scalar holomorphic functions $f_1,\dots,f_\ell:\Omega\rightarrow\mathbb{C}$, with $\ell\leq n^2$. In many applications, $T$ is naturally given in this form, usually with a small number of terms, $\ell\ll n^2$.

The linear eigenvalue problem is a particular case where $T$ can be written as $T(z)=A-zB$ with $A,B\in\mathbb{C}^{n\times n}$. Compared to the linear eigenvalue problem, the solution structure of the NEP is more diverse. A NEP may have no solution at all, finitely many solutions, or infinitely many solutions, and eigenvectors belonging to distinct eigenvalues need not be linearly independent. The survey by \citet{Guttel:2017:NEP} reviews the solution structure of NEPs in detail. Next, we highlight the main points that are relevant for the rest of the paper.

The algebraic multiplicity of an eigenvalue $\lambda$ is defined as the multiplicity of the root of $\det T(z)$ at $z=\lambda$. The geometric multiplicity of $\lambda$ is the dimension of the null space of $T(\lambda)$. An eigenvalue is semisimple if its algebraic and geometric multiplicities coincide. 

The Smith form is a factorization, analog of the Jordan form in linear problems, that reveals the structure of eigenvalues (including partial multiplicities) together with (generalized) right and left eigenvectors. The Smith form can also be used to work with the resolvent. In particular, Keldysh's theorem provides an explicit formula for the resolvent. In the simpler case when all eigenvalues $\lambda_i$ are semisimple, the expression becomes
\begin{equation}\label{eq:resolvent}
T^{-1}(z)=\sum_{i=1}^k(z-\lambda_i)^{-1}x_iy_i^*+R(z),
\end{equation}
where $k$ is the number of eigenvalues $\lambda_i$ contained in $\Omega$, $x_i$ and $y_i$ are the corresponding right and left eigenvectors, respectively, normalized so that $y_i^*T'(\lambda_i)x_i=1$, and $R(z)$ is some function holomorphic in $\Omega$.

Some algorithms rely on the concept of invariant pair \cite{Kressner:2009:BNM,Beyn:2011:CEI}. A pair $(X,H)\in\mathbb{C}^{n\times k}\times\mathbb{C}^{k\times k}$ is invariant if
\begin{equation}\label{eq:invpair}
\mathbb T(X,H):=\frac{1}{2\pi\mathrm{i}}\int_{\Gamma}T(z)X(zI-H)^{-1}dz=0,
\end{equation}
where $\Gamma$ is a contour enclosing the eigenvalues of $H$. If $T$ is expressed in the split form \eqref{eq:split}, then the definition can be written as
\begin{equation}
\mathbb T(X,H):=A_1Xf_1(H)+\dots+A_\ell Xf_\ell(H)=0,
\end{equation}
where $f_i(H)$ denotes a matrix function \cite{Higham:2008:FMT}. Furthermore, an invariant pair is called minimal if there is an integer $p$ such that $\operatorname{rank}\mathcal{V}_p(X,H)=k$ for
\begin{equation}\label{eq:invpair-split}
\mathcal{V}_p(X,H):=\left[\begin{smallmatrix}
    X\\ XH \\ \vdots \\ XH^{p-1}
\end{smallmatrix}\right].
\end{equation}
The smallest such $p$ is called the minimality index of the pair $(X,H)$. If $(X,H)$ is a minimal invariant pair for $T$, then the eigenvalues of $H$ are eigenvalues of $T$~\cite{Kressner:2009:BNM}.

The algorithms discussed in \S{\ref{sec:solvers}} compute approximate solutions of~{\eqref{eq:nep}}. The accuracy of a computed solution $(\tilde x,\tilde\lambda)$ can be assessed by means of the backward error
\begin{equation}
\eta(\tilde x,\tilde\lambda)=\min\{\varepsilon:(T(\tilde\lambda)+\Delta T(\tilde\lambda))\tilde x=0,\;\|\Delta T\|_\Omega\leq \varepsilon\|T\|_\Omega\},
\end{equation}
where $\|T\|_\Omega=\sup_{z\in\Omega}\|T(z)\|$ for some matrix norm $\|\cdot\|$ on $\mathbb{C}^{n\times n}$. When $T(\lambda)$ is provided in split form~{\eqref{eq:split}}, \citet{Guttel:2017:NEP} suggest using the explicit expression based on the scaled residual
\begin{equation}\label{eq:sressplit}
\eta(\tilde x,\tilde\lambda)=\frac{\|T(\tilde\lambda)\tilde x\|}{f(\tilde\lambda)\|\tilde x\|},\qquad\text{with}\quad f(\tilde\lambda)=\sum_{i=1}^\ell|f_i(\tilde\lambda)|\|A_i\|,
\end{equation}
which is a straightforward generalization of \cite[Theorem 1 and Lemma 3]{Tisseur:2000:BEC}.
If the scaled residual in~{\eqref{eq:sressplit}} is small then the approximate eigenpair $(\tilde x,\tilde\lambda)$ is the exact eigenpair of a nearby problem.
If the problem is not expressed in the split form, we will use the following upper bound instead,
\begin{equation}\label{eq:sres}
\eta(\tilde x,\tilde\lambda)=\frac{\|T(\tilde\lambda)\tilde x\|}{\|T(\tilde\lambda)\|\|\tilde x\|}.
\end{equation}

\section{Design of the NEP module\label{sec:design}}

In this section we describe how the \texttt{NEP} module is organized, including its relation with other SLEPc objects, and also how the user interacts with it, either with the programmatic interface or the command-line options.

Currently implemented methods are (details are postponed to \S{\ref{sec:solvers}}):
\begin{itemize}
\item Successive linear problems (SLP), a Newton-type iteration where a linear eigenproblem is solved at each step to get the eigenvalue correction.
\item Residual inverse iteration (RII), that computes a Newton-type correction of the eigenvector by multiplying $T(\sigma)^{-1}$ times the residual.
\item Nonlinear Arnoldi, that builds an expanding subspace with the vectors generated by RII, and chooses approximate eigenpairs via Rayleigh-Ritz projection.
\item Polynomial interpolation, where a matrix polynomial $P(\lambda)$ is built by evaluating $T(\cdot)$ at a few points, then a solver of class \texttt{PEP} is used to solve the polynomial eigenproblem.
\item NLEIGS, which is based on a Krylov iteration operating on a companion-type linearization of a rational interpolant of the nonlinear function.
\end{itemize}

\subsection{Overview of SLEPc\label{sec:slepc}}

\begin{figure}
\begin{tikzpicture}
\tikzstyle{solver}=[draw,fill=gray!20,font=\sffamily\bfseries]
\tikzstyle{aux}=[draw]
\node[solver] (nep) at (1,3) {\texttt{NEP}};
\node[solver] (pep) at (0,2) {\texttt{PEP}};
\node[solver] (svd) at (2.3,2) {\texttt{SVD}};
\node[solver] (eps) at (.8,1) {\texttt{EPS}};
\node[solver] (mfn) at (3.3,1) {\texttt{MFN}};
\node[aux] (st) at (-.2,0) {\texttt{ST}};
\node[aux] (bv) at (0.6,0) {\texttt{BV}};
\node[aux] (ds) at (1.4,0) {\texttt{DS}};
\node[aux] (rg) at (2.2,0) {\texttt{RG}};
\node[aux] (fn) at (3,0)   {\texttt{FN}};
\draw[-latex] (nep) -- (pep);
\draw[-latex] (nep) -- (eps);
\draw[-latex] (pep) -- (eps);
\draw[-latex] (svd) -- (eps);
\draw[-stealth,dashed] (pep) -- (st);
\draw[-stealth,dashed] (eps) -- (st);
\draw[-stealth,dashed] (nep) .. controls (1.3,1.5) and (2,1) .. (fn);
\draw[-stealth,dashed] (mfn) -- (fn);
\end{tikzpicture}
\caption{\label{fig:slepc}Components of the SLEPc library. Shaded modules represent the main solvers, and white blocks are auxiliary classes. Solid arrows indicate that a solver may sometimes use another solver of different type, and dashed arrows mean dependence on auxiliary objects for some computations or due to the user interface (arrows from almost all solvers to \texttt{BV}, \texttt{DS}, and \texttt{RG} have been omitted).}
\end{figure}
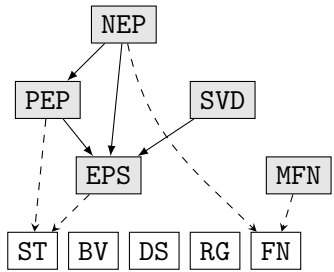

SLEPc, the Scalable Library for Eigenvalue Problem Computations \cite{Hernandez:2005:SSF,Roman:2019:SUM2}, is a parallel library intended mainly for the solution of large-scale eigenvalue problems. The latest versions also include functionality for matrix functions. Figure \ref{fig:slepc} illustrates the relation among the different SLEPc components.

The main solver class is \texttt{EPS} (Eigenvalue Problem Solver), for linear eigenvalue problems (standard or generalized). For nonlinear eigenvalue problems, there is the \texttt{PEP} module (specific for polynomial eigenvalue problems) and the \texttt{NEP} module, which is the focus of this paper. In addition, there are two more solver classes that we will not discuss further in this paper: \texttt{SVD} for singular value computations and \texttt{MFN} for computing the action of a sparse matrix function on a vector.

Figure \ref{fig:slepc} also shows five auxiliary classes. The first one (\texttt{ST}) provides functionality for spectral transformations such as shift-and-invert, which are used in linear and polynomial problems but not in the \texttt{NEP} module. The other four are relevant for this paper: \texttt{BV} for functionality relative to basis vectors, \texttt{DS} for (sequential) dense solvers used in projected problems, \texttt{RG} for specifying a region of the complex plane, and \texttt{FN} for specifying a scalar function and computing dense matrix functions.

SLEPc relies on PETSc (Portable, Extensible Toolkit for Scientific Computation \cite{Balay:2020:PUM}), so in addition to the components depicted in Figure \ref{fig:slepc}, SLEPc also depends on PETSc classes, most notably data objects such as vectors (\texttt{Vec}) and matrices (\texttt{Mat}), and solvers for linear systems of equations (\texttt{KSP}) and preconditioners (\texttt{PC}).

PETSc and SLEPc have an object-oriented design, where objects of the different classes interact with each other via a public interface, hiding the implementation details within them. All solver classes have several implementations that can be selected at run time, so for instance a \texttt{KSP} object may run GMRES, Bi-CGStab, or others, depending on the user's choice. The particular solver to be used, and many other settings, may be specified within the program as well as with command-line options.

The model of parallelism is based on the message-passing standard MPI\footnote{Local computations can exploit a second level of parallelism via a multi-threaded BLAS or with GPU computing, but we do not discuss this here.}. Every object is bound to an MPI communicator since its creation, and some of the object's operations are collective, meaning that the associated computation is carried out in parallel among all MPI processes in the communicator. This implies that some data structures must be distributed across those processes. Almost all functionality of PETSc and SLEPc is parallelized, one of the most notable exceptions being the matrix factorizations (LU, Cholesky) that are sequential in PETSc. As a workaround for this, PETSc seamlessly integrates third-party packages such as MUMPS \cite{Amestoy:2001:FAM} to perform parallel direct linear solves.

PETSc's object-oriented model is somewhat limited, as it is implemented in the C language. Some object-oriented features such as overloading or templates are not available. Object composition is very common (for instance, a \texttt{NEP} object may \emph{contain} an \texttt{EPS} object), and, regarding inheritance, the hierarchy of classes is essentially flat, with polymorphism achieved via \emph{subtyping} (for instance, a \texttt{NEP} object can be of type \texttt{slp}, \texttt{rii}, \texttt{narnoldi}, \texttt{interpol}, or \texttt{nleigs}). This subtyping mechanism involves a nomenclature convention, as will be seen when describing the user interface in \S{\ref{sec:interface}} and \S{\ref{sec:solvers}}. The function \texttt{XXXYYYSetZZZ} belongs to type \texttt{YYY} within class \texttt{XXX} and sets the attribute \texttt{ZZZ}. If \texttt{YYY} is omitted, it refers to an attribute of the base class. For example, \texttt{NEPSetTolerances} sets two attributes (\texttt{tol} and \texttt{max\_it}) that are common to all \texttt{NEP} objects, and can be set also via the command line with \texttt{-nep\_tol} and \texttt{-nep\_max\_it}, whereas \texttt{NEPRIISetHermitian} sets the hermitian flag that is specific of the \texttt{RII} type, and the syntax for setting it in the command-line is \texttt{-nep\_rii\_hermitian}. Objects can define a prefix for their associated command-line options, and this is used in the case of object composition. For example, a solver of type \texttt{SLP} contains an object of type \texttt{EPS}, whose options can be set, e.g., with \texttt{-nep\_slp\_eps\_tol}. We conclude the discussion about object-oriented features noting that PETSc provides a mechanism to extend a class by adding new user-defined types via registering the constructor of the new type. In this way, a user can implement a new \texttt{NEP} solver and register it in the system via \texttt{NEPRegister}.

\subsection{User interface\label{sec:interface}}

The basic usage of eigensolvers in SLEPc requires creating the solver object, passing the matrices that define the problem, and then invoking the ``solve'' method. After that, the solution is kept internally to the solver and can be retrieved with a few function calls. The solution process can be customized with different settings as we will discuss later.

In \texttt{EPS} and \texttt{PEP}, the user defines the problem by providing the coefficient matrices. In the case of \texttt{NEP}, the matrix-valued function $T(\cdot)$ requires also information about the nonlinear function. We will now describe how the application code can define $T(\cdot)$ as well as the derivative $T'(\cdot)$, that we will call the Jacobian and is required by some solvers. There are two ways of doing this: with callback functions and using the split form.

\begin{figure}
\begin{Verbatim}[fontsize=\footnotesize,numbers=left,numbersep=6pt,xleftmargin=30mm]
NEP         nep;       /* eigensolver context            */
Mat         T, J;      /* Function and Jacobian matrices */
Vec         xr, xi;    /* eigenvector, x                 */
PetscScalar kr, ki;    /* eigenvalue, k                  */
ApplCtx     ctx;       /* user-defined context           */
PetscInt    j, nconv;
PetscReal   error;

NEPCreate( PETSC_COMM_WORLD, &nep );
/* omitted: create and preallocate T and J matrices */
NEPSetFunction( nep, T, T, FormFunction, &ctx );
NEPSetJacobian( nep, J, FormJacobian, &ctx );
NEPSetFromOptions( nep );
NEPSolve( nep );
NEPGetConverged( nep, &nconv );
for (j=0; j<nconv; j++) {
  NEPGetEigenpair( nep, j, &kr, &ki, xr, xi );
  NEPComputeError( nep, j, NEP_ERROR_RELATIVE, &error );
}
NEPDestroy( &nep );
\end{Verbatim}
\caption{\label{fig:callback}Example C code for basic solution with \texttt{NEP} using callbacks.}
\end{figure}

\subsubsection{Callbacks}\label{sec:callback}
We start describing the callback function mechanism. Figure \ref{fig:callback} shows basic code for solving a NEP in this way. Two variables are used to represent both the eigenvalue and the eigenvector, and the rationale for this has to do with real versus complex arithmetic as discussed below (\S\ref{sec:complex}). By default, the solver attempts to compute just one eigenpair, but this and other settings can be adjusted (\S\ref{sec:settings}). After calling the solver, the example checks the number of converged eigenpairs, and extracts each of them from the solver object. It also computes the relative residual error, $\|T(\lambda_j)x_j\|/|\lambda_j|$.

Lines 11-12 of the example code are the relevant ones for the callback function mechanism. With \texttt{NEPSetFunction} the user provides a PETSc matrix \texttt{T}, a user-defined function \texttt{FormFunction}, and an optional context \texttt{ctx}. The user-defined function (not shown in Figure \ref{fig:callback}) must have the prototype
\begin{Verbatim}[fontsize=\footnotesize,xleftmargin=10mm]
PetscErrorCode FormFunction(NEP nep,PetscScalar lambda,Mat T,Mat P,void *ctx)
\end{Verbatim}
and must contain code that computes the entries and assembles the matrix $T(\lambda)$ from the value of $\lambda$ (argument \texttt{lambda}) and possibly some application-specific parameters stored in the data structure pointed by \texttt{ctx}. The computed matrix is the argument \texttt{T} (note that this matrix is created only once in the main program and reused here). Similarly, the user-defined function \texttt{FormJacobian} builds a PETSc matrix \texttt{J} that stores $T'(\lambda)$ for a given $\lambda$. The main difference between the two callback functions is that \texttt{FormFunction} has an additional argument \texttt{P} intended to build a preconditioner matrix (that for basic usage is equal to matrix \texttt{T}).

With this scheme, whenever the SLEPc solver needs to evaluate $T(\lambda)$ or $T'(\lambda)$, it invokes the user subroutines to have this done. As we will see in \S\ref{sec:solvers}, most solvers rely on evaluating these matrices (or just $T(\lambda)$ in some cases) for different values of the parameter. Often, it is also necessary to perform linear system solves with $T(\sigma)$, usually for a fixed value $\sigma$ referred to as the target (\S\ref{sec:settings}).

We point out that it is also possible to solve the problem in a matrix-free fashion, that is, just implementing subroutines that compute the action of $T(\lambda)$ or $T'(\lambda)$ on a vector, rather than explicitly building these two matrices. This can be done by means of PETSc \emph{shell matrices}.

\subsubsection{Split form}\label{sec:split}
An alternative way to specify the nonlinear operator $T(\cdot)$ is the split form~\eqref{eq:split}, where the user provides a separate set of constant matrices and nonlinear functions. This is the recommended usage because most problems coming from real applications can be written naturally in the split form with only a few terms, and also because, compared to the callback mechanism, it is simpler (the user does not have to worry about how to compute the derivative) and more robust (no chance to introduce bugs in user callbacks). Still, the callback mechanism is required for problems that cannot be described easily in the split form.

To illustrate the use of the split form in SLEPc, let us focus on a particular NEP coming from the discretization of a parabolic partial differential equation with time delay $\tau$,
\begin{equation}\label{eq:delay}
(-\lambda I + A + \mathrm{e}^{-\tau\lambda}B)x = 0.
\end{equation}
In this case, the matrix-valued function $T(\lambda)$ has a polynomial part ($-\lambda I + A$) and a nonlinear part ($\mathrm{e}^{-\tau\lambda}B$). In SLEPc we do not distinguish between the polynomial and nonlinear parts, so all terms are treated equally. Hence, for this example we need to build three matrices (the identity $I$, $A$, and $B$) and three analytic functions
\begin{equation}\label{eq:delayfuns}
f_1(\lambda)=-\lambda,\quad
f_2(\lambda)=1,\quad
f_3(\lambda)=\mathrm{e}^{-\tau\lambda}.
\end{equation}
Figure~\ref{fig:split} illustrates how \texttt{NEPSetSplitOperator} is used to provide an array of matrices and functions. The code for creating the matrices is not shown in the figure. The functions \eqref{eq:delayfuns} are managed with SLEPc's auxiliary object \texttt{FN}, that can be used to represent mathematical functions from a set of predefined types and combinations thereof, in particular:
\begin{itemize}
  \item Polynomial and rational functions, by providing the list of coefficients of the numerator and (optionally) the denominator.
  \item The (principal) square root, $x^\frac{1}{2}$, and the inverse square root, $x^{-\frac{1}{2}}$.
  \item Transcendental functions: the exponential, the logarithm, and the $\varphi$-functions defined as
  \begin{equation}
  \varphi_0(x)=\mathrm{e}^x,\qquad \varphi_1(x)=\frac{\mathrm{e}^x-1}{x},\qquad \varphi_k(x)=\frac{\varphi_{k-1}(x)-1/(k-1)!}{x}.
  \end{equation}
  \item Scaling factors can be specified for both the argument and the function, with \texttt{FNSetScale}, to form, e.g., $g(x)=\beta \mathrm{e}^{\alpha x}$.
  \item The special type \texttt{FNCOMBINE} allows creating new functions by combining two \texttt{FN} objects with either addition, multiplication, division, or function composition.
  \item If the previous functionality is not enough to cover the needs of an application, the user can extend SLEPc by adding a new \texttt{FN} function implementation via \texttt{FNRegister}.
\end{itemize}

\begin{figure}
\begin{Verbatim}[fontsize=\footnotesize,numbers=left,numbersep=6pt,xleftmargin=30mm]
  FNCreate(PETSC_COMM_WORLD,&f1);  /* f1 = -lambda */
  FNSetType(f1,FNRATIONAL);
  coeffs[0] = -1.0; coeffs[1] = 0.0;
  FNRationalSetNumerator(f1,2,coeffs);

  FNCreate(PETSC_COMM_WORLD,&f2);  /* f2 = 1 */
  FNSetType(f2,FNRATIONAL);
  coeffs[0] = 1.0;
  FNRationalSetNumerator(f2,1,coeffs);

  FNCreate(PETSC_COMM_WORLD,&f3);  /* f3 = exp(-tau*lambda) */
  FNSetType(f3,FNEXP);
  FNSetScale(f3,-tau,1.0);

  mats[0] = A;  funs[0] = f2;
  mats[1] = Id; funs[1] = f1;
  mats[2] = B;  funs[2] = f3;
  NEPSetSplitOperator(nep,3,mats,funs,SUBSET_NONZERO_PATTERN);
\end{Verbatim}
\caption{\label{fig:split}Excerpt of C code illustrating the definition of the nonlinear function in split form.}
\end{figure}

Every \texttt{FN} object, representing a function $f:\mathbb{C}\rightarrow\mathbb{C}$, provides the \texttt{FNEvaluateFunction} and \texttt{FNEvaluateDerivative} operations to evaluate $f(x)$ and $f'(x)$ for a given value $x$. When the nonlinear operator of the NEP is represented in split form~\eqref{eq:split}, SLEPc is able to build matrix $T(\lambda)$ by evaluating $f_i(\lambda)$, $i=1,\dots,\ell$, together with $\ell-1$ matrix \emph{axpy} operations, and $T'(\lambda)$ can be built similarly by evaluating $f_i'(\lambda)$ instead. Note that the \emph{axpy} operations for sparse matrices as implemented in PETSc can be optimized in the case that the sparsity patterns are related; this is hinted with the last argument of \texttt{NEPSetSplitOperator} (see line 18 of Figure~\ref{fig:split}).

We note in passing that \texttt{FN} is also capable of evaluating matrix functions $f:\mathbb{C}^{k\times k}\rightarrow\mathbb{C}^{k\times k}$~\cite{Higham:2008:FMT}. This is required by some methods, as will be pointed out in \S\ref{sec:solvers}.

\subsubsection{Real and complex arithmetic}\label{sec:complex}
A feature of PETSc (and hence of SLEPc) is that the library is built for either real or complex scalars. This implies that the basic type \texttt{PetscScalar} (and consequently all data structures such as matrices and vectors) is either a real number or a complex number. It is not possible to mix real and complex scalars in the same program\footnote{This design decision is sometimes criticized as a weakness of PETSc, but it eliminates many complications/inefficiencies.}. This has an impact in SLEPc's user interface, because real matrices may have complex eigenpairs.

In linear eigenproblem (\texttt{EPS}), the solver typically works internally with a (partial) real or complex Schur form, in real and complex arithmetic, respectively. When the final computed solution is to be returned to the user, it is represented as follows. In real arithmetic, two \texttt{PetscScalar} variables are used for the eigenvalue, storing the real and imaginary parts, and similarly two \texttt{Vec} objects are employed for the eigenvector. If the eigenvalue is real, the imaginary part is set to zero. In complex arithmetic, only one variable is used for the eigenvalue and one vector for the eigenvector.

This strategy carries over to nonlinear eigenproblems (\texttt{PEP} and \texttt{NEP}). However, the \texttt{NEP} case needs further clarification because, as described above, the solvers need to evaluate $T(\lambda)$ (or its derivative) and the question is what happens if $\lambda$ is complex when the computation is done in real arithmetic. This case would oblige to represent $T(\lambda)$ with two \texttt{Mat} objects, one for the real part and one for the imaginary part, with the consequent complication of the user interface: for instance the user callback \texttt{FormFunction} would require two scalar arguments \texttt{lambdar} and \texttt{lambdai}, and four matrix arguments \texttt{Tr}, \texttt{Ti}, \texttt{Pr}, \texttt{Pi}. Furthermore, this design would involve a high complication in real arithmetic to use PETSc's linear solvers in \texttt{KSP}.

To avoid the complication described above, we decided to forbid the evaluation of $T(\lambda)$ for a complex $\lambda$ in case of real arithmetic. If this happens, the solver would fail with an error message. Even though this may seem a severe limitation, in practice this implies that solvers will not be able to compute complex eigenvalues in real arithmetic, and for real eigenvalues the search region (described next) must be restricted to an interval of the real line.

\subsubsection{Settings}\label{sec:settings}
We now describe general settings of the \texttt{NEP} object. The user can change them within the code or with a command-line option. For instance, calling \texttt{NEPSetTarget(nep,1.4)} is equivalent to running the program adding the command-line argument \texttt{-nep\_target 1.4} (command-line options are processed in the \texttt{NEPSetFromOptions} operation, see line 13 of Figure~\ref{fig:callback}).

Basic settings include the following:
\begin{itemize}
  \item \texttt{NEPSetDimensions} allows specifying the number of eigenvalues to compute (\texttt{nev}) and the maximum allowed dimension of the search subspace (\texttt{ncv}, only relevant in some solvers).
  \item \texttt{NEPSetTolerances} indicates the tolerance for the stopping criterion and the maximum number of iterations.
  \item \texttt{NEPSetWhichEigenpairs} is used to indicate which eigenvalues are of interest, e.g., those with largest magnitude, with largest real part, or closest to a given target value $\sigma$.
  \item \texttt{NEPSetTarget} is used to provide the value of the target, $\sigma$.
  \item \texttt{NEPSetType} selects the solver to be used. Available solvers are described in \S\ref{sec:solvers}.
  \item \texttt{NEPSetProblemType} can be used to indicate that the problem is \texttt{rational} (as opposed to \texttt{general}). The aim of problem types in SLEPc is to allow users to give a hint to the solver so that it can make optimizations in some cases.
\end{itemize}

As other eigensolver classes, \texttt{NEP} can be used with a region defined via an auxiliary object \texttt{RG}. The \texttt{RG} object is a simple mechanism in SLEPc to define regions of the complex plane, such as an interval, a polygon, an ellipse, or a ring. In \texttt{NEP}, the meaning of the region is different depending on the solver type (see details in \S\ref{sec:solvers}):
\begin{itemize}
  \item In interpolation methods, the region is used to select the interpolation nodes, and also as a filtering mechanism (eigenvalue approximations outside the region are discarded).
  \item In the other methods (Newton-based) the region is not taken into account.
\end{itemize}

\subsubsection{Left eigenvectors and the resolvent}\label{sec:leftvecs}
In order to compute left eigenvectors~\eqref{eq:nep-left}, the user must select a two-sided variant of the solver with \texttt{NEPSetTwoSided}. Currently, only the NLEIGS solver (\S\ref{sec:nleigs}) supports this option. Then, after the solver has finished, left eigenvectors can be retrieved with \texttt{NEPGetLeftEigenvector}. Furthermore, the operation \texttt{NEPApplyResolvent} is available to compute the action of the resolvent~\eqref{eq:resolvent} on a vector.

\section{Solvers\label{sec:solvers}}

This section describes the solvers that are available in the \texttt{NEP} module. The solvers are very different from each other and it is convenient that the user knows how they work in order to decide which one is the most appropriate for a particular problem. In general, NLEIGS (\S\ref{sec:nleigs}) is often our most competitive solver, but other solvers may be better suited for special cases.

\subsection{Classical Iterations\label{sec:classical}}

\begin{algorithm}[t]
  \KwIn{Initial eigenvalue approximation $\lambda^{(0)}$}
  \KwOut{Computed eigenpair $(x^{(k)},\lambda^{(k)})$}
  \For{k=0,1,2,\dots}{
    Evaluate $A=T(\lambda^{(k)}),\quad B=T'(\lambda^{(k)})$ \;
    \lIf{$k>0$ {\rm and} $\eta(x^{(k)},\lambda^{(k)})<\texttt{tol}$}{exit}\label{alg:slp:conv}
    Compute $(x^{(k+1)},\mu^{(k)})$, the smallest magnitude eigenpair of $Ax=\mu Bx$\label{alg:slp:eig} \;
    Update $\lambda^{(k+1)}=\lambda^{(k)}-\mu^{(k)}$ \;
  }
\caption{Successive Linear Problems (SLP)}\label{alg:slp}
\end{algorithm}

\citet{Ruhe:1973:ANE} proposed the method of successive linear problems (SLP). This method is based on linearization via Taylor's formula,
\begin{equation}\label{eq:taylor}
T(\lambda+\mu)=T(\lambda)+\mu T'(\lambda)+\frac{\mu^2}{2}R(\lambda,\mu),
\end{equation}
from which high-order terms $R$ are discarded. This results in an iteration where at each step the correction $\mu$ to the eigenvalue is obtained by computing the smallest magnitude eigenvalue of a linear eigenvalue problem, see Algorithm~\ref{alg:slp}. Several comments are in order:
\begin{itemize}
  \item The convergence criterion shown in line~\ref{alg:slp:conv} of Algorithm~\ref{alg:slp} is based on the scaled residual $\eta(x,\lambda)$ defined in~{\eqref{eq:sres}} and~{\eqref{eq:sressplit}}, but the user can also choose the absolute residual, the relative residual, or even a custom convergence test.
  \item The eigenproblem in line~\ref{alg:slp:eig} is solved via an \texttt{EPS} solver. In particular, it is solved with shift-and-invert Krylov-Schur with target $\sigma=0$, which in practice is equivalent to computing largest magnitude eigenvalues of $A^{-1}Bx=\mu^{-1}x$. Therefore, matrix $T(\lambda^{(k)})$ is factorized at each step $k$.
  \item When solving the eigenproblem in line~\ref{alg:slp:eig}, $x^{(k)}$ is set as initial guess for the wanted eigenvector, $x^{(k+1)}$.
  \item Algorithm~\ref{alg:slp} computes just one eigenpair. Computation of several eigenpairs requires adding a deflation mechanism, described in \S\ref{sec:deflation}.
\end{itemize}

The main drawback of SLP is that a factorization of $T(\lambda^{(k)})$ is required at each iteration. The same problem is present in nonlinear inverse iteration, also discussed in \cite{Ruhe:1973:ANE}. We have implemented the residual inverse iteration (RII), a modification proposed by \citet{Neumaier:1985:RII} that replaces $T(\lambda^{(k)})$ with $T(\sigma)$, for a constant value $\sigma$, with the consequent reduction of computational cost. To derive these methods, start with Newton's method applied to the system of $n+1$ nonlinear equations consisting of $T(\lambda)x=0$ together with a normalization condition. The Newton iteration can be written as
\begin{equation}\label{eq:newton}
T(\lambda^{(k)})x^{(k+1)}=-(\lambda^{(k+1)}-\lambda^{(k)})T'(\lambda^{(k)})x^{(k)},
\end{equation}
followed by normalization of $x^{(k+1)}$. The Newton update \eqref{eq:newton} can be expressed in the form of nonlinear inverse iteration, where the new eigenvector approximation is computed by solving the linear system
\begin{equation}\label{eq:nonii}
T(\lambda^{(k)})x^{(k+1)}=T'(\lambda^{(k)})x^{(k)},
\end{equation}
and then the eigenvalue approximation $\lambda^{(k+1)}$ is updated with a correction. Neumaier's idea is to rewrite \eqref{eq:newton} as
\begin{equation}\label{eq:neumaier}
\begin{split}
x^{(k)}-x^{(k+1)}&=x^{(k)}+(\lambda^{(k+1)}-\lambda^{(k)})T(\lambda^{(k)})^{-1}T'(\lambda^{(k)})x^{(k)}\\
&=T(\lambda^{(k)})^{-1}\left(T(\lambda^{(k)})+(\lambda^{(k+1)}-\lambda^{(k)})T'(\lambda^{(k)})\right)x^{(k)}.
\end{split}
\end{equation}
The expression in parenthesis is a first-order approximation of $T(\lambda^{(k+1)})$, as in \eqref{eq:taylor}, resulting in the iteration
\begin{equation}\label{eq:rii}
x^{(k+1)}=x^{(k)}-T(\lambda^{(k)})^{-1}T(\lambda^{(k+1)})x^{(k)}.
\end{equation}
The main difference of residual inverse iteration \eqref{eq:rii}, with respect to Newton, is that the new eigenvalue approximation $\lambda^{(k+1)}$ must be computed in advance. The algorithm in \cite{Neumaier:1985:RII} replaces $\lambda^{(k)}$ with a fixed shift $\sigma$, showing that convergence is not affected.

\begin{algorithm}[t]
  \KwIn{Target value $\sigma$, initial guess $x^{(0)}$}
  \KwOut{Computed eigenpair $(x^{(k)},\lambda^{(k+1)})$}
  $\lambda^{(0)}=\sigma$ \;
  \For{k=0,1,2,\dots}{
    $\lambda^{(k+1)}=\lambda^{(k)}$\label{alg:rii:scalare1} \;
    \Repeat{$|\mu|/|\lambda^{(k+1)}|<\sqrt{\varepsilon}$\label{alg:rii:scalare2}}{
      $\lambda^{(k+1)}=\lambda^{(k+1)}-\mu$, with
      $\;\mu = \frac{{x^{(k)}}^*T(\sigma)^{-1}T(\lambda^{(k+1)})x^{(k)}}{{x^{(k)}}^*T(\sigma)^{-1}T'(\lambda^{(k+1)})x^{(k)}}$ \;
    }
    Compute residual $r=T(\lambda^{(k+1)})x^{(k)}$\label{alg:rii:resid}\;
    \lIf{$\eta(x^{(k)},\lambda^{(k+1)})<\texttt{tol}$}{exit}
    \lIf{required}{update $\sigma$, evaluate $T(\sigma)$}\label{alg:rii:update}
    Solve $T(\sigma)v=r$ \label{alg:rii:solve} \;
    Update and normalize $x^{(k+1)}=\tilde x^{(k+1)}/\|\tilde x^{(k+1)}\|$, with $\tilde x^{(k+1)}=x^{(k)}-v$ \;
  }
\caption{Residual Inverse Iteration (RII)}\label{alg:rii}
\end{algorithm}

Our implementation is shown in Algorithm \ref{alg:rii}. Lines \ref{alg:rii:scalare1}--\ref{alg:rii:scalare2} represent a Newton iteration to solve the scalar nonlinear equation
\begin{equation}\label{eq:rii-scalar}
{x^{(k)}}^*T(\sigma)^{-1}T(z)x^{(k)}=0,
\end{equation}
where the computed solution is taken as $\lambda^{(k+1)}$. As suggested by Neumaier, this equation can be replaced with the cheaper version ${x^{(k)}}^*T(z)x^{(k)}=0$ if $T(\lambda)$ is Hermitian. In our implementation, the user can choose between the two versions with \texttt{NEPRIISetHermitian}. The stopping criterion for this loop uses a tolerance equal to the square root of the machine epsilon, $\varepsilon$. In line \ref{alg:rii:update}, we give the user the possibility of updating $\sigma$ with the value of $\lambda^{(k+1)}$. The user interface function \texttt{NEPRIISetLagPreconditioner} allows to do this, by specifying the number of iterations after which the preconditioner $T(\sigma)$ must be updated (0 means a fixed $\sigma$, and 1 implies updating at every iteration as in Newton). The linear solve of line \ref{alg:rii:solve} is carried out via a PETSc's \texttt{KSP} object, that can also be customized by the user (by default, GMRES with block Jacobi is used). In case of iterative linear solves, we offer the possibility of using a constant tolerance or an exponentially decreasing tolerance (\texttt{NEPRIISetConstCorrectionTol}).

\subsection{Nonlinear Arnoldi\label{sec:narnoldi}}

\begin{algorithm}[t]
  \KwIn{Target value $\sigma$, initial guess $x^{(0)}$}
  \KwOut{Computed eigenpair $(x^{(k)},\lambda^{(k)})$}
  Initialize basis $V_1 = [x^{(0)}/\|x^{(0)}\|]$ \;
  \For{k=1,2,\dots}{
    Compute $(y,\tilde\lambda)$ satisfying projected problem $V_k^*T(\tilde\lambda)V_ky=0$ \label{alg:narnoldi:proj} \;
    Compute Ritz pair $(x^{(k)}=V_ky,\lambda^{(k)}=\tilde\lambda)$ and residual $r=T(\lambda^{(k)})x^{(k)}$ \;
    \lIf{$\eta(x^{(k)},\lambda^{(k)})<\texttt{tol}$}{exit}
    Solve $T(\sigma)v = r$ \label{alg:narnoldi:solvecorr} \;
    Expand subspace: $\tilde v=v-V_kV_k^*v,\quad V_{k+1}=[V_k,\tilde v/\|\tilde v\|]$ \label{alg:narnoldi:expand} \;
  }
\caption{Nonlinear Arnoldi (N-Arnoldi)}\label{alg:narnoldi}
\end{algorithm}

Both the SLP and RII methods presented in the previous section are single-vector iterations. However, in the area of linear eigenproblems it is well-known that single-vector iterations can be less effective than algorithms based on projection onto an expanding subspace. This idea can also be applied to the NEP. \citet{Voss:2004:AMN} suggests a projection method with RII iterates, that is, correction vectors obtained in step~\ref{alg:rii:solve} of Algorithm~\ref{alg:rii} are used to expand a subspace whose orthogonal basis is $V_k$. Then, instead of solving a one-dimensional nonlinear equation~\eqref{eq:rii-scalar}, the problem is projected onto this $k$-dimensional subspace, resulting in a NEP of size $k$
\begin{equation}\label{eq:narnoldi}
V_k^*T(\tilde\lambda)V_ky=0.
\end{equation}
An eigenvalue $\tilde\lambda$ satisfying~\eqref{eq:narnoldi} is the Ritz approximate eigenvalue $\lambda^{(k)}$ of the original NEP, and the corresponding eigenvector $y$ is related to the Ritz vector $x^{(k)}=V_ky$. Voss called this method Nonlinear Arnoldi (N-Arnoldi). Algorithm~\ref{alg:narnoldi} sketches our implementation of this method in SLEPc.

Next, we discuss how the N-Arnoldi solver in SLEPc manages the projected problem~\eqref{eq:narnoldi}. With the split form, \eqref{eq:narnoldi} turns into
\begin{equation}\label{eq:narnoldi-split}
\left(\sum_{i=1}^{\ell}B_if_i(\tilde\lambda)\right)y=0,\quad B_i=V_k^*A_iV_k.
\end{equation}
The projected matrices $B_i$ are computed incrementally, that is, their $k$th rows and columns are computed at step $k$ using the last column of $V_k$. The projected NEP \eqref{eq:narnoldi-split} is solved with \texttt{DS}, which is one of SLEPc's auxiliary objects depicted in Figure~\ref{fig:slepc}. \texttt{DS} stands for Direct Solver (or Dense System), and is intended to solve small-sized eigenproblems whose coefficient matrices are dense. For linear eigenproblems, \texttt{DS} involves calling one or more LAPACK subroutines. But for NEP dense systems, we need to provide a nonlinear eigensolver at the \texttt{DS} level. Currently, we have implemented a dense version of the SLP method of Algorithm~\ref{alg:slp}. This dense implementation receives the $B_i$ matrices and a handle to the $f_i$ functions, and performs all steps of the algorithm with dense computations sequentially, with LAPACK's \texttt{\_ggev} to compute the eigenvalue correction. 

\subsection{Deflation\label{sec:deflation}}

The methods presented so far (SLP, RII and N-Arnoldi) approximate just one eigenpair. However, SLEPc's implementation of these methods support the computation of more than one eigenpair. Subsequent eigenpairs are computed by implementing a deflation mechanism, to avoid unwanted convergence to previously computed eigenpairs. Deflation can be achieved via a non-equivalence transformation to the matrix-valued function $T(\cdot)$ that maps already computed eigenvalues to infinity~\cite{Guttel:2017:NEP}. We have opted for implementing the deflation technique developed by \citet{Effenberger:2013:RSC}, as described below.

Effenberger's deflation is based on the concept of minimal invariant pair described in \S\ref{sec:math}. Suppose a minimal invariant pair $(X,H)\in\mathbb{C}^{n\times k}\times\mathbb{C}^{k\times k}$ of the NEP has already been obtained, with minimality index $p$. We want to compute an extended pair
\begin{equation}\label{eq:invpextend}
(\tilde X,\tilde H)=\left(\begin{bmatrix}X&x\end{bmatrix},
\begin{bmatrix}H&t\\0&\lambda\end{bmatrix}\right)
\end{equation}
such that it contains a new eigenpair $(x,\lambda)$ and is also a minimal invariant pair of the same NEP. As we will see below, imposing the conditions of invariant pair and minimality on the extended pair~\eqref{eq:invpextend} results in an extended nonlinear eigenvalue problem (of size $n\!+\!k$),
\begin{equation}\label{eq:sistext}
\tilde T(\lambda)\begin{bmatrix}x\\t\end{bmatrix}=\begin{bmatrix}T(\lambda)&U(\lambda)\\A(\lambda)&B(\lambda)\end{bmatrix}\begin{bmatrix}x\\t\end{bmatrix}=0,
\end{equation}
where $U(\lambda)$, $A(\lambda)$ and $B(\lambda)$ are matrix-valued functions of dimensions $n\times k$, $k\times n$ and $k\times k$, respectively. Solving the extended NEP~\eqref{eq:sistext} provides the required data ($\lambda, x, t$) for the wanted extension~\eqref{eq:invpextend}. Hence, in SLEPc the implementations of SLP, RII and N-Arnoldi apply the respective algorithm to the extended NEP~\eqref{eq:sistext}, operating by blocks as discussed below, to compute one eigenpair at a time.

To enforce the definition of invariant pair~\eqref{eq:invpair} on the extended pair, Effenberger uses the fact that
\begin{equation}
(zI-\tilde H)^{-1}=\begin{bmatrix}(zI-H)^{-1} & (zI-H)^{-1}t(z-\lambda)^{-1}\\ 0 & (z-\lambda)^{-1}\end{bmatrix},
\end{equation}
and writes
\begin{equation}
\mathbb T(\tilde X,\tilde H)=\begin{bmatrix}\mathbb T(X,H) & T(\lambda)x+U(\lambda)t\end{bmatrix},
\end{equation}
with
\begin{equation}
U(\lambda)=\frac{1}{2\pi\mathrm{i}}\int_{\Gamma}T(z)X(zI-H)^{-1}(z-\lambda)^{-1}dz,
\end{equation}
where the contour $\Gamma$ encloses the eigenvalues of $H$ and $\lambda$. Since $(X,H)$ is invariant, the condition $\mathbb T(\tilde X,\tilde H)=0$ holds if and only if $T(\lambda)x+U(\lambda)t=0$. This relation constitutes the first block equation of the extended NEP~\eqref{eq:sistext}. The second block equation, $A(\lambda)x+B(\lambda)t=0$, results from imposing the condition that $(\tilde X,\tilde H)$ is minimal, \eqref{eq:invpair-split}, with minimality index not exceeding $p\!+\!1$. \citet{Effenberger:2013:RSC} proves this and gives the expressions
\begin{equation}
A(\lambda)=\sum_{i=0}^{p}\lambda^i(XH^i)^*,\quad B(\lambda)=\sum_{i=1}^{p}(XH^i)^*Xq_i(\lambda),
\end{equation}
with
\begin{equation}
q_i(\lambda)=\sum_{j=0}^{i-1}\lambda^jH^{i-j-1}.
\end{equation}
$A(\lambda)$ and $B(\lambda)$ are matrix polynomials, which are easy to evaluate. However, evaluating $U(\lambda)$ is more involved. It can be shown~\cite{Effenberger:2013:RSC} that if $\lambda$ is not an eigenvalue of $H$,
\begin{equation}\label{eq:ushort}
U(\lambda)=T(\lambda)X(\lambda I-H)^{-1}.
\end{equation}
Alternatively, if $T(\cdot)$ is expressed in the split form~\eqref{eq:split}, then
\begin{equation}\label{eq:usplit}
U(\lambda)=\sum_{i=1}^{\ell}A_iX\phi_i(\lambda),
\end{equation}
where $\phi_i(\lambda)$ is the $k\times k$ leading principal submatrix of the matrix function $f_i\left(\left[\begin{smallmatrix}H & I\\0 & \lambda I\end{smallmatrix}\right]\right)$. For this, our solvers make use of matrix functions as implemented in the \texttt{FN} module, as was mentioned at the end of \S\ref{sec:split}. If any of the provided $f_i$ functions lacks a matrix function implementation, we fall back to~{\eqref{eq:ushort}} hoping that $\lambda$ is not an eigenvalue of $H$.

We now discuss how we have implemented the main operations in a block fashion, that is, without explicitly building the operator of the extended NEP~\eqref{eq:sistext}. According to Algorithms~\ref{alg:slp},~\ref{alg:rii} and~\ref{alg:narnoldi}, we need to perform matrix-vector multiplications with $\tilde T(\lambda)$ and $\tilde T'(\lambda)$, and linear solves with $\tilde T(\sigma)$. The general strategy is that we have \texttt{Vec} objects representing vectors $\left[\begin{smallmatrix}x\\t\end{smallmatrix}\right]$ of length $n\!+\!k$, from which we can easily extract two subvectors $x$ and $t$, also \texttt{Vec} objects, of length $n$ and $k$, respectively, to perform the operation by blocks, the first vector being a parallel vector (distributed across all processes) while the second one is sequential (stored redundantly in all processes to minimize communication, see \S\ref{sec:parallel}).

Matrix-vector products with the extended operator, $\left[\begin{smallmatrix}y_1\\y_2\end{smallmatrix}\right]=\tilde T(\lambda)\left[\begin{smallmatrix}z_1\\z_2\end{smallmatrix}\right]$, can be accomplished with
\begin{equation}\label{eq:block-mv}
\begin{split}
  y_1=&\,T(\lambda)z_1+\sum_{i=1}^\ell A_iX\phi_i(\lambda)z_2,\\
  y_2=&\sum_{i=0}^{p}\lambda^i(XH^i)^*z_1+\sum_{i=1}^{p}(XH^i)^*Xq_i(\lambda)z_2.
\end{split}
\end{equation}
For the linear system solves
$\left[\begin{smallmatrix}T(\sigma)&U(\sigma)\\A(\sigma)&B(\sigma)\end{smallmatrix}\right]\left[\begin{smallmatrix}x_1\\x_2\end{smallmatrix}\right]=\left[\begin{smallmatrix}b_1\\b_2\end{smallmatrix}\right]$
we define the Schur complement
\begin{equation}
  S(\sigma):=B(\sigma)-A(\sigma)T(\sigma)^{-1}U(\sigma),
\end{equation}
that in the case of callback representation is computed as $B(\sigma)-A(\sigma)X(\sigma I-H)^{-1}$ using~\eqref{eq:ushort} and assuming that $\sigma$ is not an eigenvalue of $H$. The operation is divided in three steps:
\begin{equation}\label{eq:block-solve}
  v=T(\sigma)^{-1}b_1\quad\left\{
  \begin{split}
  x_2&=S(\sigma)^{-1}(b_2-A(\sigma)v),\\
  x_1&=v-T(\sigma)^{-1}U(\sigma)x_2=v-X(\sigma I-H)^{-1}x_2,
  \end{split}
  \right.
\end{equation}
where we have used~\eqref{eq:ushort} again for $x_1$. The computation of $v$ amounts to a standard linear solve of size $n$ with PETSc's \texttt{KSP}.

Finally, N-Arnoldi requires an additional operation (step~\ref{alg:narnoldi:proj} of Algorithm~\ref{alg:narnoldi}) to perform the projection of the extended operator onto a basis of length $n\!+\!k$,
\begin{equation}\label{eq:defl-proj}
  \begin{split}
  \begin{bmatrix}V_1^*&V_2^*\end{bmatrix}\begin{bmatrix}T(\lambda)&U(\lambda)\\A(\lambda)&B(\lambda)\end{bmatrix}\begin{bmatrix}V_1\\V_2\end{bmatrix}=
  V_1^*T(\lambda)V_1+
  V_1^*U(\lambda)V_2+
  V_2^*A(\lambda)V_1+
  V_2^*B(\lambda)V_2 =\\
  \sum_{i=1}^{\ell}V_1^*A_iV_1f_i(\lambda)
  +\sum_{i=1}^{\ell}V_1^*A_iX\phi_i(\lambda)V_2+
  \sum_{i=0}^{p}\lambda^iV_2^*(XH^i)^*V_1
  +\sum_{i=1}^{p}V_2^*(XH^i)^*Xq_i(\lambda)V_2.
  \end{split}
\end{equation}
The first term contains the $B_i$ matrices of~\eqref{eq:narnoldi-split}, that are updated for each new basis vector.

In some problems, using the deflation may produce slightly inaccurate eigenpairs. The reason is that deflating with respect to an eigenpair that has been computed with a certain error (given by the user's tolerance) introduces a perturbation that limits the attainable accuracy of subsequent eigenpairs. This means that, for instance, the second eigenpair has about one significant digit less than the first one. As a remedy, we have introduced a threshold parameter in SLP and RII (\texttt{NEPRIISetDeflationThreshold}) that works as follows. At the beginning, the solver operates with the extended operator so that previous eigenpairs are deflated. When the relative residual error of the current eigenpair is below the threshold, the solver switches to the simpler variant without deflation. In that way, the last part of the computation proceeds without the perturbation, and better accuracy can be attained. We assume that once the threshold has been reached, the iterates will be already in the convergence basin of the current eigenpair. Still, there is risk of misconvergence if this option is used. In the experiments of \S\ref{sec:performance}, this option was needed only in one problem (turned off by default).

\subsection{Polynomial Interpolation\label{sec:interpol}}

The methods presented in \S\ref{sec:classical} and \S\ref{sec:narnoldi} derive from Newton's method in one way or another. A completely different strategy for solving the NEP consists in approximating $T(\lambda)$ with a matrix polynomial $P(\lambda)$ (or a rational matrix-valued function $R(\lambda)$, see \S\ref{sec:nleigs}) and then computing the solution of this alternative eigenproblem, hoping that its solution is in good agreement with the solution of the original problem.

In this section we discuss SLEPc's \texttt{interpol} solver, that implements the method proposed by \citet{Effenberger:2012:CIN} based on a Chebyshev interpolation of $T(\lambda)$. This solver can be used for the particular case when the wanted eigenvalues of the NEP are real and lie within a prescribed closed interval of the real line, $\mathcal{I}=[a,b]$. The user specifies the interval with an \texttt{RG} object of type \texttt{interval}. The method has two stages. First, it computes the coefficient matrices of $P_d(\lambda)$, the Chebyshev polynomial (of the first kind) of degree $d$ that interpolates $T(\lambda)$ at the Chebyshev nodes
\begin{equation}\label{eq:interpolnu}
\nu_i:=\cos \left(\left(i+\frac{1}{2}\right)\frac{\pi}{d+1}\right),\quad i=0,\dots,d,
\end{equation}
in the interval $[-1,1]$. The mapping from $\mathcal{I}$ to $[-1,1]$ usually requires a change of variable,
\begin{equation}
\lambda=\frac{b-a}{2}\theta+\frac{b+a}{2}.
\end{equation}
If $P_d$ is expressed in the basis of Chebyshev polynomials $\tau_k(\cdot)$,
\begin{equation}\label{eq:polinterp}
P_d(\lambda)=\frac{C_0}{2}\tau_0(\lambda)+\sum_{k=1}^{d}C_k\tau_k(\lambda),
\end{equation}
then $C_k=\frac{2}{d+1}\sum_{i=0}^{d}T(\nu_i)\tau_k(\nu_i)$, for $\nu_i$ defined in~\eqref{eq:interpolnu},
or equivalently,
\begin{equation}\label{eq:polinterpcoef}
C_k=\frac{2}{d+1}\sum_{i=0}^{d}T(\cos \omega_i)\cos(k\omega_i),\quad \omega_i=\left(i+\frac{1}{2}\right)\frac{\pi}{d+1}.
\end{equation}
Secondly, the polynomial eigenvalue problem $P_d(\lambda)x=0$ is solved and the eigensolutions $(x,\lambda)$ are returned as solutions of the NEP. The nice thing of this approach is that the polynomial eigenproblem is managed with a SLEPc solver of class \texttt{PEP}, so here we leverage all the functionality available in that module. By default, the polynomial eigenproblem is solved via linearization and a memory-efficient Krylov solver. The details of how this linearization is done when the polynomial is expressed in the Chebyshev basis can be found in~\cite{Campos:2016:PKS}. An alternative to linearization is to employ a polynomial Jacobi-Davidson solver with support for Chebyshev basis~\cite{Campos:2020:PJD}. In any case, the solver must be used in a combination with a region filter (\texttt{RG}) in order to discard computed eigenvalue approximations that do not lie in the interval (or its immediate vicinity).

\citet{Effenberger:2012:CIN} show that the approximation error $\|T(\lambda)-P_d(\lambda)\|_F$ decreases exponentially, uniformly in $[-1,1]$, when the interpolation degree is increased. They also show that, for sufficiently large $d$, the eigenvalues obtained in the associated polynomial eigenproblem are good approximations to eigenvalues of the NEP whenever those are close to the interval $\mathcal{I}$. The degree of the polynomial, $d$, is a user setting that can be specified at run time (\texttt{NEPInterpolSetInterpolation}). It must be noted that the solver will compute the eigenpairs of the polynomial eigenproblem up to the accuracy requested by the user (tolerance), but the actual error corresponding to the NEP may be larger if the value $d$ is not large enough. There is no automatic procedure to choose an appropriate value of $d$.

\subsection{Rational Interpolation\label{sec:nleigs}}

As mentioned in the previous section, the polynomial interpolation solver is restricted to the case of real eigenvalues. Another limitation is that the approximation error gets worse if the interpolation points are not sufficiently far away from the singularities of $T(\cdot)$. In such cases it is preferable to have recourse to a rational interpolation that takes into account the singularities. In SLEPc we provide a solver based on the NLEIGS method~\cite{Guttel:2014:NCF}, consisting in a (rational) Krylov method operating on the linearization pencil resulting from the rational interpolation. We provide a memory-efficient implementation in the sense that a compact (tensor-product) representation of the Krylov basis is employed throughout the computation (we call it the TOAR representation by analogy with the polynomial eigensolvers of~\cite{Campos:2016:PKS}). In this way, our solver fits the framework of the CORK method~\cite{Beeumen:2015:CRK}. Furthermore, we have enriched the solver with other features such as automatic determination of singularities and a two-sided variant to compute left eigenvectors. All the details are described next.

\subsubsection{Overview of NLEIGS}

Following the notation of~\cite{Guttel:2014:NCF}, the goal is to find eigenvalues located in a compact target set $\Sigma\subset\Omega$, in which $T(\cdot)$ is analytic, by approximating $T$ with a rational matrix-valued function whose poles are selected from the set of singularities of $T$, denoted by $\Xi$. Considering the degree-graded rational Newton basis functions defined by the recursion
\begin{equation}\label{eq:nleigsbj}
b_0(\lambda)=1,\quad b_{j}(\lambda)=\frac{\lambda-\sigma_{j-1}}{\beta_{j}(1-\lambda/\xi_{j})}b_{j-1}(\lambda),\quad j=1,2,\dots
\end{equation}
with nonzero poles at $\xi_j\in\Xi$, the rational matrix
\begin{equation}\label{eq:nleigsq}
R_d(\lambda):=\sum_{j=0}^{d}b_j(\lambda)D_j
\end{equation}
is built in such a way that it interpolates $T$ at the nodes $\sigma_j\in\partial\Sigma$ (the boundary of $\Sigma$). The scaling factors $\beta_j$ in~\eqref{eq:nleigsbj} are chosen so that $\max_{\lambda\in\partial\Sigma}|b_j(\lambda)|=1$.
If all interpolation nodes $\sigma_j$ are pairwise distinct, the coefficient matrices $D_j$ of~\eqref{eq:nleigsq} can be computed easily, from the interpolation conditions $R_j(\sigma_j)=T(\sigma_j)$, by means of the recurrence
\begin{equation}\label{eq:nleigsdd}
D_0=\beta_0T(\sigma_0),\quad
D_{j}=\frac{T(\sigma_j)-R_{j-1}(\sigma_j)}{b_j(\sigma_j)},\quad j=1,2,\dots.
\end{equation}
These matrices are referred to as rational divided difference matrices.
In case $T$ is expressed in the split form~\eqref{eq:split}, they can be written as
\begin{equation}\label{eq:nleigsdsplit}
D_j=\sum_{i=0}^{\ell}d_{i}^jA_i,\quad j\ge 0,
\end{equation}
where $d_{i}^j$ denotes the $j$th rational divided difference corresponding to the scalar function $f_i$. As shown in~\cite{Guttel:2014:NCF}, these scalar divided differences can be obtained in a numerically stable way using matrix functions, through the expression
\begin{equation}\label{eq:nleigsfdd}
d_{i}^j=\beta_0e_j^Tf_i(H_{d+1}K_{d+1}^{-1})e_1,
\end{equation}
where $H_{d+1}$ and $K_{d+1}$ are bidiagonal matrices given by
\begin{equation}
H_{d+1}:=\left[\begin{matrix}
\sigma_0&&&\\\beta_1&\sigma_1&&\\&\ddots&\ddots&\\&&\beta_d&\sigma_d
\end{matrix}\right],\quad
K_{d+1}:=\left[\begin{matrix}
1&&&\\\beta_1/\xi_1&1&&\\&\ddots&\ddots&\\&&\beta_d/\xi_d&1
\end{matrix}\right].
\end{equation}
In SLEPc's implementation of NLEIGS, the $D_j$ matrices are computed via~\eqref{eq:nleigsdd} if the NEP is represented with callback functions, and using the expression~\eqref{eq:nleigsfdd} if the split form~\eqref{eq:split} is being used. The latter case relies on the matrix function evaluation capabilities of \texttt{FN}, as discussed in \S\ref{sec:split}. Furthermore, in the split case the $D_j$ matrices~\eqref{eq:nleigsdsplit} are not computed explicitly, and instead the solver works with them implicitly, that is, operating directly with the $A_i$ matrices that appear in the definition of operator $T$.

The degree $d$ of the interpolant~\eqref{eq:nleigsq} is determined at run time, by evaluating the norms of the rational divided difference matrices as they are generated, until the relation $\|D_d\|/\|D_0\|<\mathtt{tol}$ holds for a given tolerance \texttt{tol}. The user can provide this tolerance and the maximum allowed degree with \texttt{NEPNLEIGSSetInterpolation}. In problems with split representation, for which the $D_j$ matrices are not available explicitly, we use the estimates of their norms proposed in~\cite{Guttel:2014:NCF}, that make use of the divided differences of the $f_i$ functions: $\delta^d:=\max\{|d_i^d|:i=0,\dots,\ell\}$.

The interpolation nodes and poles that determine the approximation $R_d(\lambda)$ of~\eqref{eq:nleigsq} are obtained as a sequence of Leja--Bagby points for $(\Sigma,\Xi)$~\cite{Guttel:2014:NCF}: starting from an arbitrary $\sigma_0\in\partial\Sigma$, select nodes $\sigma_j\in\partial\Sigma$ and poles $\xi_j\in\Xi$ recursively such that they satisfy the conditions
\begin{equation}\label{eq:nleigsleja}
\max_{z\in\partial\Sigma} |s_j(z)|=|s_j(\sigma_{j+1})|\quad\text{and}\quad\min_{z\in\Xi} |s_j(z)|=|s_j(\xi_{j+1})|,
\end{equation}
for
\begin{equation}
s_j(\lambda):=\frac{\prod_{k=0}^{j}(\lambda-\sigma_k)}{\prod_{k=1}^{j}(1-\lambda/\xi_k)},\quad j=0,1,\dots.
\end{equation}
The fact that the interpolation points $\sigma_j$ lie at the boundary $\partial\Sigma$ is a direct consequence of the maximum module theorem.
In our solver, the target set $\Sigma$ is specified by the user via a region object (\texttt{RG}, \S\ref{sec:settings}), and hence its boundary $\partial\Sigma$ is discretized automatically using the functionality offered by \texttt{RG}. In contrast, defining the singularity set $\Xi$ is more subtle and we offer several possibilities:
\begin{itemize}
  \item The user may provide a callback function (with \texttt{NEPNLEIGSSetSingularitiesFunction}) that returns a sequence of points as a discretization of $\Xi$. This function may return less points than requested by the method, in which case the remaining values are set to infinity. In the extreme case that all  $\xi_j=\infty$, then the basis functions of~\eqref{eq:nleigsbj} become polynomials and a Newton polynomial interpolation at nodes $\sigma_j$ is effectively used.
  \item If the callback function is not provided, but the NEP is of type \texttt{rational} (\S\ref{sec:settings}) and has been defined in the split form~\eqref{eq:split} using only rational functions or a simple combination (sum or product) of quotients of polynomials, then the solver is able to automatically determine the poles. The poles of a quotient of polynomials are obtained by computing the roots of the denominator via the eigenvalues of the associated companion matrix.
  \item Otherwise, a discretization of $\Xi$ is computed via the AAA algorithm~\cite{Nakatsukasa:2018:AAR}. In particular, we follow the procedure developed by \citet{Elsworth:2019:CBB}. Note that the AAA algorithm is used only to obtain the $\xi_j$ points, not to approximate the NEP.
\end{itemize}

It is worth mentioning the special case when $T(\cdot)$ is a rational matrix-valued function of type $(p,q)$. Then the interpolant $R_d$ of~\eqref{eq:nleigsq} with $d=\max\{p,q\}$ will be exact for any choice of the sampling points $\sigma_j$. As a consequence, our implementation will be able to solve rational eigenvalue problems (REP) via linearization in an alternative way compared to the method of \citet{Su:2011:SRE}, with the advantage that the REP does not need to be expressed in any specific form.

\medskip
Once the rational approximation~\eqref{eq:nleigsq} has been built, a linearization is carried out resulting in a linear eigenvalue problem
\begin{equation}\label{eq:nleigslin}
\Alin y=\lambda \Blin y,
\end{equation}
whose eigenvalues $\lambda$ are the same as for the rational eigenproblem $R_d(\lambda)x=0$ and whose eigenvectors have the form
\begin{equation}
y=\begin{bmatrix}
b_0(\lambda)x\\\vdots\\b_{d-1}(\lambda)x
\end{bmatrix}.
\end{equation}
For the matrices $\Alin$ and $\Blin$ of the linearization~\eqref{eq:nleigslin}, we use the simplification proposed in~\cite{Guttel:2014:NCF} of setting the last pole $\xi_d=\infty$, resulting in
\begin{equation}\label{eq:nleigsmatlin}
  \Alin=\begin{bmatrix}
    D_0       & D_1      & \dots  & D_{d-2}       & (D_{d-1}-\frac{\sigma_{d-1}}{\beta_d}D_d) \\
    \sigma_0I & \beta_1I &        &               & \\
              & \ddots   & \ddots &               & \\
              &          & \ddots & \beta_{d-2}I  & \\
              &          &        & \sigma_{d-2}I & \beta_{d-1}I
  \end{bmatrix},\quad
  \Blin=\begin{bmatrix}
    0 & 0                      & \dots  & 0                              &-\frac{D_d}{\beta_d} \\
    I & \frac{\beta_1}{\xi_1}I &        &                                & \\
      & \ddots                 & \ddots &                                & \\
      &                        & \ddots & \frac{\beta_{d-2}}{\xi_{d-2}}I & \\
      &                        &        & I                              & \frac{\beta_{d-1}}{\xi_{d-1}}I
  \end{bmatrix}.
\end{equation}
To solve the linear eigenproblem~\eqref{eq:nleigslin}, \citet{Guttel:2014:NCF} propose two main variants of the rational Krylov method~\cite{Ruhe:1984:RKS}. In the dynamic variant, the approximation and the linearization are built incrementally as the Krylov subspace grows, with rational Krylov shifts equal to the interpolation nodes $\sigma_j$, while in the static variant the linearization is created a priori and then solved. In SLEPc we have implemented the static variant only, where the user can optionally provide the list of rational Krylov shifts (\texttt{NEPNLEIGSSetRKShifts}), otherwise only one shift is used (the target), which is equivalent to the shift-and-invert Krylov--Schur method~\cite{Stewart:2001:KAL}. In our experience, the rational Krylov method tends to be less efficient for large problems, since it requires several matrix factorizations. We provide details of the Krylov--Schur case in the next subsections.

\subsubsection{Expansion of the Krylov basis}\label{sec:nleigsexp}

In our NLEIGS solver, the linearization matrices are never built explicitly, and instead the Krylov--Schur method proceeds by exploiting their block structure.
To generate a new Krylov vector, we need to multiply the last vector of the basis with matrix
\begin{equation}\label{eq:nleigssinvert}
\Slin=(\Alin-\sigma \Blin)^{-1}\Blin,
\end{equation}
where $\Alin$ and $\Blin$ are given in~\eqref{eq:nleigsmatlin}, and $\sigma\in\Sigma$ is the shift or target value (a value around which the eigenvalues are sought). To derive the recurrences that will apply matrix $\Slin$ implicitly, we consider the factorization $(\Alin-\sigma \Blin)\Pi=U_\sigma L_\sigma$, where $\Pi=\left[\begin{smallmatrix} 0&I_{(d-1)n}\\I_n&0 \end{smallmatrix}\right]$ is a permutation matrix, and
\begin{equation}
L_\sigma\!=\!\begin{bmatrix}
  \frac{1}{b_{d-1}(\sigma)}I &  &  & \\
  -\frac{b_0(\sigma)}{b_{d-1}(\sigma)}I &I &  & \\
  \vdots &  &\ddots & \\
  -\frac{b_{d-2}(\sigma)}{b_{d-1}(\sigma)}I &  &  & I
\end{bmatrix}\!,\quad
U_\sigma\!=\!\begin{bmatrix}
  R_d(\sigma) &D_0 &D_1      &\dots      &D_{d-2}     \\
  &(\sigma_0\!-\!\sigma)I      &\beta_1(1\!-\!\frac{\sigma}{\xi_1})I &                                 &                                 \\
  &                            &(\sigma_1\!-\!\sigma)I           &\ddots &                                 \\
  &                            &                                 &\ddots              &\beta_{d-2}(1\!-\!\frac{\sigma}{\xi_{d-2}})I \\
  &                            &                                 &                                 &(\sigma_{d-2}\!-\!\sigma)I
\end{bmatrix}\!.
\end{equation}
Using this factorization, the product $w=\Slin x=\Pi L_\sigma^{-1}U_\sigma^{-1}\Blin x$ is carried out by first solving the block upper triangular system $U_\sigma y=\Blin x$, by means of the recurrence
\begin{equation}\label{eq:nleigstoarreca}
\begin{split}
y^{d-1}&=\frac{1}{\sigma_{d-2}-\sigma}x^{d-2}+\frac{\beta_{d-1}}{(\sigma_{d-2}-\sigma)\xi_{d-1}}x^{d-1},\\
y^j&=\frac{1}{\sigma_{j-1}-\sigma}x^{j-1}+\frac{\beta_{j}}{(\sigma_{j-1}-\sigma)\xi_j}x^{j}-\frac{\beta_j}{\sigma_{j-1}-\sigma}\left(1-\frac{\sigma}{\xi_j}\right)y^{j+1},\quad j=d\!-\!2,\dots,1,\\
y^0&=R_d(\sigma)^{-1}\left(-D_0y^1-D_1y^2-\dots-D_{d-2}y^{d-1}-\frac{1}{\beta_d}D_dx^{d-1}\right).
\end{split}
\end{equation}
In the above expressions, $x^i$ denotes the $i$th block of vector $x=\operatorname{vec}(x^0,\dots,x^{d-1})$.
Secondly, the block lower triangular system $L_\sigma\tilde w=y$ is solved with
\begin{equation}\label{eq:nleigstoarrecb}
\begin{split}
\tilde w^0&=b_{d-1}(\sigma)y^0,\\
\tilde w^j&=y^j+b_{j-1}(\sigma)y^0,\quad j=1,\dots,d-1.
\end{split}
\end{equation}
Finally, the solution vector is obtained by applying the permutation $w=\Pi\tilde w$.

The operations in~\eqref{eq:nleigstoarreca}--\eqref{eq:nleigstoarrecb} involve vector \emph{axpy} operations, matrix-vector products with the divided difference matrices, and the construction of $R_d(\sigma)$ and its inverse. Of course, the inverse is not built explicitly, but instead a (sparse) linear solve is done (using a \texttt{KSP} linear solver object) which usually implies a factorization. It is noteworthy that, in our implementation, when the NEP has been defined in the split form~\eqref{eq:split}, the last expression of~\eqref{eq:nleigstoarreca} operates with the divided difference matrices $D_j$ without forming them explicitly. Since the $D_j$ consist in linear combinations of the problem matrices $A_i$, those operations are rearranged in a way that they are expressed in terms of the $A_i$ matrices only.

\subsubsection{Krylov basis representation}\label{sec:nleigstoar}

We provide two different implementations of NLEIGS, one that operates with explicitly stored Krylov vectors (full basis) and another one with a compact representation (TOAR basis). The advantage of the full basis is that it is possible to use any of the \texttt{EPS} eigensolvers (not only Krylov--Schur) to solve the linear eigenvalue problem~\eqref{eq:nleigslin}, whereas the TOAR basis allows a significant reduction of both computational and storage cost. By default, the compact representation is used, and the other one can be selected with the user option \texttt{NEPNLEIGSSetFullBasis}.

We now describe the specialized version of Krylov--Schur that handles the compact representation of the Krylov basis, in a similar way as the TOAR solver for polynomial eigenvalue problems in \texttt{PEP}~\cite{Campos:2016:PKS}. The idea is to exploit the block structure of the matrices $\Alin$ and $\Blin$ of the linearization, and derive dependency relations among the $d$ blocks of the generated Krylov vectors, whose global dimension is $dn$. In this way, it is possible to define a basis $U_{k+d}$ of vectors of length $n$, from which all blocks of the Krylov basis $V_k$ can be reconstructed, resulting in the relation
\begin{equation}\label{eq:toarrep}
V_k=(I_d\otimes U_{k+d})G_k,
\end{equation}
for some matrix of coefficients $G_k$.

The Krylov--Schur method applied to the linear eigenproblem~\eqref{eq:nleigslin} produces an Arnoldi relation of order $k$ for matrix $\Slin$~\eqref{eq:nleigssinvert}, that can be written as
\begin{equation}\label{eq:arnoldi}
\Blin V_k=(\Alin-\sigma \Blin)(V_kH_k+h_{k+1,k}v_{k+1}e_k^*).
\end{equation}
Considering that the Krylov basis $V_k$ is divided in $d$ blocks of $n$ consecutive rows, $\{V_k^i\}_{i=0}^{d-1}$, and equating the last $d\!-\!1$ block-rows of~\eqref{eq:arnoldi}, we get $d\!-\!1$ relations that reveal the linear dependency between the vectors of the set formed by the columns of blocks $\{V_k^i\}_{i=0}^{d-1}$ and the vectors $\{v_{k+1}^i\}_{i=0}^{d-1}$. This implies the equivalence of subspaces
\begin{equation}\label{eq:blockspand}
\operatorname{span}(\cup_{i=0}^{d-1}\{V_k^ie_j\}_{j=1}^{k}\cup\{v_{k+1}^i\}_{i=0}^{d-1})=\operatorname{span}(\{V_k^0e_j\}_{j=1}^{k}\cup\{v_{k+1}^i\}_{i=0}^{d-1}),
\end{equation}
meaning that each of the $d$ parts of the $k\!+\!1$ Arnoldi vectors stored in the columns of $V_{k+1}$ can be obtained from linear combinations of $k\!+\!d$ vectors, that is, $V_k$ can be expressed as
\begin{equation}\label{eq:toarblocksd}
\begin{bmatrix}
V_{k}^i&v_{k+1}^i
\end{bmatrix}=U_{k+d}\begin{bmatrix}
G_{k}^i&g_{k+1}^i
\end{bmatrix}\,\quad i=0,\dots,d-1,
\end{equation}
where $U_{k+d}\in\mathbb{C}^{n\times (k+d)}$, $\{G_k^i\}_{i=0}^{d-1}\subset\mathbb{C}^{(k+d)\times k}$ and $\{g_{k+1}^i\}_{i=0}^{d-1}\subset\mathbb{C}^{k+d}$. Both $U_{k+d}$ and $G_k$ must have orthonormal columns by construction. This is the TOAR representation of~\eqref{eq:toarrep}, that has been described in~\cite{Campos:2016:PKS} in the context of polynomial eigenvalue problems.

The recurrences \eqref{eq:nleigstoarreca}--\eqref{eq:nleigstoarrecb} that generate a new Arnoldi vector must be adapted to operate with vectors in the TOAR representation. Assuming that the current Arnoldi basis is~\eqref{eq:toarblocksd}, the function
\begin{equation}\label{eq:nleigsoperator}
[u_{k+d+1},g] = \mathtt{expand}(\{D_i\}_{i=0}^{d},U_{k+d},g_{k+1}),
\end{equation}
generates a vector $u_{k+d+1}$ to extend the TOAR basis (if necessary), and the coefficients $g$ used to represent the expansion vector $w=\Slin v_{k+1}$ in terms of such basis by the following steps. Firstly, we compute the TOAR coefficients of the $y^j$ vectors in~\eqref{eq:nleigstoarreca},
\begin{equation}\label{eq:nleigs-toar1}
\left\{
\begin{aligned}
\tilde g^{d-1}&=\frac{1}{\sigma_{d-2}-\sigma}g^{d-2}_{k+1}+\frac{\beta_{d-1}}{(\sigma_{d-2}-\sigma)\xi_{d-1}}g^{d-1}_{k+1},\\
\tilde g^j&=
\frac{1}{\sigma_{j-1}-\sigma}g_{k+1}^{j-1}+\frac{\beta_{j}}{(\sigma_{j-1}-\sigma)\xi_j}g_{k+1}^{j}-\frac{\beta_j}{\sigma_{j-1}-\sigma}\left(1-\frac{\sigma}{\xi_j}\right)\tilde g^{j+1},\quad j=d\!-\!2,\dots,1.\\
\end{aligned}\right.
\end{equation}
To compute $y^0$, it is necessary to reconstruct the preceding vectors using the expressions $y^i=U_{k+d}\tilde g^i$, for $i=1,\dots,d\!-\!1$, and the last block of the input vector as $v_{k+1}^{d-1}=U_{k+d}g_{k+1}^{d-1}$. Then
\begin{equation}\label{eq:nleigs-toar2}
y^0=R_d(\sigma)^{-1}\left(-D_0y^1-D_1y^2-\dots-D_{d-2}y^{d-1}-\frac{1}{\beta_d}D_dv_{k+1}^{d-1}\right),
\end{equation}
which is used to generate vector $u_{k+d+1}$ that extends the orthogonal basis $U_{k+d}$ (unless it is linearly dependent). Then, from the orthogonalization coefficients, we obtain the coordinates $\tilde g^0$ that express $y^0$ with respect to the extended basis,
\begin{equation}\label{eq:nleigs-toar3}
y^0=U_{k+d+1}\tilde g^0.
\end{equation}
Finally, the coefficients $g^i$ of~\eqref{eq:nleigsoperator} can be obtained from the coefficients $\tilde g$ using~\eqref{eq:nleigstoarrecb} and taking into account the permutation $\Pi$, as
\begin{equation}\label{eq:nleigs-toar4}
\left\{
\begin{aligned}
g^j&=\tilde g^{j+1}+b_{j}(\sigma)\tilde g^0,\quad j=0,\dots,d-2,\\
g^{d-1}&=b_{d-1}(\sigma)\tilde g^0.
\end{aligned}\right.
\end{equation}

\begin{algorithm}[t]
\KwIn{Divided difference matrices $\{D_i\}_{i=0}^d\subset \mathbb{C}^{n\times n}$, vectors $\{w^i\}_{i=0}^{d-1}\subset\mathbb{C}^{n}$ ($d$ parts of initial vector), number of iterations $k\in\mathbb{N}$}
\KwOut{${H}_k\in \mathbb{C}^{k\times k}$, $h_{k+1,k}\in\mathbb{R}$, $U_{k+d}\in \mathbb{C}^{n\times (k+d)}$, $\{G_k^i\}_{i=0}^d\subset\mathbb{C}^{(k+d)\times k}$, $\{g_{k+1}^i\}_{i=0}^{d-1}\subset\mathbb{C}^{k+d}$ satisfying the Arnoldi relation in TOAR representation}
$G_0\leftarrow [\quad]$, $H_0\leftarrow [\quad]$ \;
$[Q,R]=\operatorname{qr}([w^0,\dots,w^{d-1}],0)$ \;
$U_d\leftarrow Q$\;
$g_1\leftarrow \operatorname{vec}R/\|\operatorname{vec}R\,\|$\;
\For{$j=1,\ldots,k$}{
$[u_{j+d},g] = \mathtt{expand}(\{D_i\}_{i=0}^{d},U_{j-1+d},g_{j})$ \label{alg:nleigs:expand}\;
$U_{j+d}\leftarrow \left[\begin{matrix}U_{j-1+d}&u_{j+d}\end{matrix}\right]$\;
$G_{j}^i\leftarrow \left[\begin{smallmatrix}G_{j-1}^i&g_{j}^i\\ 0&0\end{smallmatrix}\right]$, $i=0,\dots,d-1$ \label{alg:nleigs:orth1}\;
$h_{j}=G_{j}^*g,\quad \hat g\leftarrow g-G_{j}h_{j}$ \label{alg:nleigs:orth2}\;
$h_{j+1,j} =\|\hat{g}\|,\quad g_{j+1}\leftarrow\hat{g}/h_{j+1,j}$ \label{alg:nleigs:norm}\;
$H_{j}\leftarrow \left[\begin{array}{c|c}\begin{smallmatrix}H_{j-1}\\0\end{smallmatrix}&h_j\end{array}\right]$\;
}
\caption{Arnoldi for NLEIGS with TOAR representation}\label{alg:nleigs}
\end{algorithm}

Algorithm~\ref{alg:nleigs} presents the Arnoldi method that is used by default within the NLEIGS solver, where step~\ref{alg:nleigs:expand} performs the expansion of the Krylov subspace as explained above. The remaining part (Gram-Schmidt orthogonalization in lines~\ref{alg:nleigs:orth1}--\ref{alg:nleigs:orth2} and normalization in line~\ref{alg:nleigs:norm}) are done in exactly the same way as in the polynomial eigensolver, as detailed in~\cite{Campos:2016:PKS}. On the other hand, we have implemented the Krylov--Schur restart on top of Algorithm~\ref{alg:nleigs}, including locking of converged eigenvalues. To keep the presentation short, we again direct interested readers to~\cite{Campos:2016:PKS} where relevant details can be found.

\subsubsection{Two-sided variant\label{sec:twosided}}

The NLEIGS solver in SLEPc provides a two-sided variant that allows computing left eigenvectors when enabled as explained in~\S\ref{sec:leftvecs}. For this, we implement a method that is similar to the two-sided Krylov--Schur of \citet{Zwaan:2017:KRT}.

The two-sided variant works with two Krylov bases, to approximate right and left eigenvectors, respectively. The left basis is built from matrix-vector products with the conjugate transpose, $\Slin^*$. Hence, we need to complement~\eqref{eq:nleigstoarreca}--\eqref{eq:nleigstoarrecb} with analogue recurrences that perform this operation by blocks. To obtain $w=\Slin^*x=\Blin^*U_\sigma^{-*}L_\sigma^{-*}\Pi^*x$ we proceed as follows. Let $y=L_\sigma^{-*}\Pi^*x$, then $L_\sigma^*y=\Pi^*x$ and we have
\begin{equation}
  \left\{
    \begin{split}
      y^i &= x^{i-1},\quad i=1,\dots d-1,\\
      y^{0} &= \bar{b}_{d-1}(\sigma)x^{d-1}+\sum_{i=1}^{d-1}\bar b_{i-1}(\sigma)y^i=\bar b_{d-1}(\sigma)x^{d-1}+\sum_{i=0}^{d-2}\bar b_{i}(\sigma)x^{i},
    \end{split}
  \right.
\end{equation}
where the bar denotes complex conjugation. For $z=U_\sigma^{-*}y$ the following relations hold,
\begin{equation}
  \begin{split}
    R_d(\sigma)^*z^0&=y^0,\\
    D_0^*z^0+(\bar \sigma_0-\bar \sigma)z^1&=y^1,\\
    D_1^*z^0+\beta_1\left(1-\frac{\bar \sigma}{\bar \xi_1}\right)z^1+(\bar\sigma_1-\bar\sigma)z^2&=y^2,\\
    &\dots\\
    D_{d-2}^*z^0+\beta_{d-2}\left(1-\frac{\bar\sigma}{\bar\xi_{d-2}}\right)z^{d-2}+(\bar\sigma_{d-2}-\bar\sigma)z^{d-1}&=y^{d-1},\\
  \end{split}
\end{equation}
from which we obtain the recurrence
\begin{equation}
  \left\{
    \begin{split}
      z^0 &= R_d(\sigma)^{-*}y^0,\\
      z^{1} &= \frac{1}{\bar\sigma_0-\bar\sigma}\left(y^1-D_0^*z^0\right),\\
      z^i &= \frac{1}{\bar\sigma_{i-1}-\bar\sigma}\left(y^i-D_{i-1}^*z^0-\beta_{i-1}\left(1-\frac{\bar\sigma}{\bar\xi_{i-1}}\right)z^{i-1}\right),\quad i=2,\dots,d-1.
    \end{split}
  \right.
\end{equation}
Finally, $w=\Blin^*z$ is computed with
\begin{equation}
  \left\{
    \begin{split}
      w^0 &= z^1,\\
      w^{i} &=\frac{\beta_i}{\bar{\xi_i}}z^i+z^{i+1},\quad i=1,\dots,d-2,\\
      w^{d-1} &= -\frac{1}{\beta_d}D_d^*z^{0}+\frac{\beta_{d-1}}{\bar\xi_{d-1}}z^{d-1} .
    \end{split}
  \right.
\end{equation}
This computation can be carried out efficiently, the most computationally expensive step being the application of $R_d(\sigma)^{-*}$, which implies a linear solve with the same \texttt{KSP} object as in~\eqref{eq:nleigstoarreca} (the factorization is reused).

Unfortunately, the above recurrences cannot be adapted to the TOAR representation, as was done for the right Krylov basis in~\eqref{eq:nleigs-toar1}--\eqref{eq:nleigs-toar4}. The reason is that the linearization~\eqref{eq:nleigslin} allows the compact representation of right eigenvectors only. There exist alternative linearizations that are appropriate for representing both right and left eigenvectors in a compact way~\cite{Lietaert:2018:CTK}, and this will be a topic of future research. In our current implementation, we restrict the two-sided variant to the case of using the full basis setting (\texttt{NEPNLEIGSSetFullBasis}, see \S\ref{sec:nleigstoar}). When this option is enabled, the solver is more expensive in terms of memory and computation, but it allows to generate the left basis required in the two-sided variant. In this case, there is no need to implement a specialized Arnoldi (Algorithm~\ref{alg:nleigs}), so we solve the linear eigenproblem using an \texttt{EPS} object (which implements two-sided Krylov--Schur).

\section{Parallel Implementation\label{sec:parallel}}

The model of parallelism of PETSc and SLEPc has been briefly introduced in \S\ref{sec:slepc}. We now provide a few details about how the algorithms described in preceding sections have been implemented in parallel. The way to proceed is very similar to other SLEPc solvers.

The parallelization is based on distributing large-scale data objects such as matrices and vectors across the available MPI processes, whereas small-scale objects such as the Hessenberg matrix $H_j$ in Algorithm~\ref{alg:nleigs} are normally stored redundantly (all processes hold a copy).

Matrices in PETSc are usually distributed by blocks of rows, that is, every process owns a contiguous range of rows. Vectors are distributed in the same way, with global and local dimensions being consistent for all objects involved in an algorithm. For example, to evaluate the residual as in step~\ref{alg:rii:resid} of Algorithm~\ref{alg:rii}, $r=T(\lambda)x$, in the case that $T$ is given in the split form~\eqref{eq:split}, the computations will be done as $r=\sum_{i=1}^{\ell}f_i(\lambda)A_ix$, which implies $\ell$ matrix-vector products and $\ell-1$ vector \emph{axpy} operations, all of them performed in parallel provided that all matrices and vectors have a compatible distribution.

A key operation in SLP (Algorithm~\ref{alg:slp}) is the explicit evaluation of the function and its derivative, $T(\lambda)$ and $T'(\lambda)$, for some scalar value $\lambda$. When working in split form, both operations are very similar and consist in $\ell-1$ matrix \emph{axpy} operations, $A=A+\alpha B$. This operation, which is implemented in PETSc, is trivially parallelizable, and its main complication is the correct management of the sparsity pattern of the involved matrices (in the \texttt{NEPSetSplitOperator} function mentioned in~\S\ref{sec:split} the user indicates whether the sparsity pattern of all $A_i$ matrices is equal or not). On the other hand, if $T$ has been specified via the callback mechanism of \S\ref{sec:callback}, then it is the user's responsibility to provide functions that compute $T(\lambda)$ and $T'(\lambda)$ in parallel.

The most costly step in SLP is the computation of the correction (step~\ref{alg:slp:eig} of Algorithm~\ref{alg:slp}) via the linear eigenproblem $A^{-1}Bx=\mu^{-1}x$. This is done with an \texttt{EPS} solver, whose parallelization involves various operations that appear in \texttt{NEP} as well and we discuss below: matrix-vector product (including linear solve), vector orthogonalization, and projected problem.

Parallel matrix-vector products involve neighbour-wise communication, and hence they should have good scalability (assuming that the problem matrices $A_i$ have an appropriate sparse pattern, e.g., coming from a mesh-based discretization). On the other hand, orthogonalization of vectors (e.g., step~\ref{alg:narnoldi:expand} of Algorithm~\ref{alg:narnoldi}) and other operations such as inner products and norms require global communication, which may limit the scalability to large number of processes. In SLEPc, we strive to minimize the impact of global communication~\cite{Hernandez:2007:PAE}.

Often it is necessary to solve linear systems of equations at each eigensolver iteration, for instance in the linear eigensolve within SLP, or in RII (step~\ref{alg:rii:solve} of Algorithm~\ref{alg:rii}), N-Arnoldi and NLEIGS. As pointed out previously, this is handled via a PETSc's \texttt{KSP} linear solver. In terms of parallelization, we repeat here that parallel factorization methods are obtained by means of external packages such as MUMPS. Note that direct linear solvers usually offer better robustness, compared to iterative ones, but may not scale well to many MPI processes.

We now discuss the solution of the projected problem, that is, the part of the algorithm that operates with small-scale matrices such as step~{\ref{alg:narnoldi:proj}} of Algorithm~{\ref{alg:narnoldi}}. As mentioned before, these small matrices are stored redundantly and the associated computation is also redundant (all processes perform the computation sequentially and independently of each other). Since this part of the algorithm is sequential, its cost must not be too high, otherwise parallel efficiency will deteriorate. In normal usage, this cost is negligible because the size of the projected problem is very small, but for the case of large number of requested eigenvalues it could become larger, so we include a user-defined parameter in \texttt{NEPSetDimensions} to keep the dimension of the projected problem bounded.

In the implementation of NLEIGS, all relevant operations have been mentioned already. The only additional comment is that the coefficient matrices $G_j$ of the compact basis are stored redundantly, so the second-level orthogonalization (steps~\ref{alg:nleigs:orth2}-\ref{alg:nleigs:norm} of Algorithm~\ref{alg:nleigs}) is also a redundant (sequential) operation.

We conclude this section stating that the most difficult operations in terms of parallel implementation are those associated with the deflation technique described in \S\ref{sec:deflation}. We have opted for increasing the memory usage to store auxiliary quantities whenever this leads to a reduction of the required MPI communication in  the main operations~\eqref{eq:block-mv},~\eqref{eq:block-solve} and~\eqref{eq:defl-proj}.
For example, in the split variant, we maintain data structures for storing $A_iX$, or $X^*X$, that are updated every time an eigenpair has converged. In this way, the matrix-vector product~\eqref{eq:block-mv} only needs to perform communication among the processes in the operations $T(\lambda)z_1$ and $X^*z_1$ when computing $y_1$ and $y_2$, respectively (recall that $z_2$ is stored redundantly in all processes). Similarly, we compute $T(\sigma)^{-1}X$ every time $X$ is extended (or $\sigma$ updated) to reduce the amount of communication in linear solves with the extended operator $\tilde T$.

\section{Performance Evaluation\label{sec:performance}}

We now present results of some computational experiments to assess the performance of the solvers in terms of accuracy, convergence, efficiency, and scalability. The executions are carried out on the Tirant III computer, which consists of 336 computing nodes, each of them with two Intel Xeon SandyBridge E5-2670 processors (16 cores each) running at 2.6 GHz with 32 GB of memory, linked with an Infiniband network. We allocated 4 MPI processes per node at most. The presented results correspond to SLEPc version 3.13, together with PETSc 3.13 and MUMPS 5.2.1. All software has been compiled with Intel C and Fortran compilers (version 18) and Intel MPI.

Table~\ref{tab:nepproblems} lists the problems used in the experiments, summarizing some properties and parameters. Here is a short description of the problems:
\begin{itemize}
\item The \textsf{delay} problem arises from the discretization of a parabolic partial differential equation with time delay $\tau$ \cite{Jarlebring:2008:SDE}, with $T(\lambda)=-\lambda I + A + \mathrm{e}^{-\tau\lambda}B$. The nonlinear character comes from the exponential, so in this case there are no singularities ($\Xi=\{\infty\}$). For the time delay parameter we have used the value $\tau=0.001$.
  \item The \textsf{loaded\_string} problem is a rational eigenproblem from a discretization of a boundary value problem describing the vibration of a string with a load of mass $m$ attached by an elastic spring of stiffness $\kappa$ \cite{Solovev:2006:PIM}. Here, the rational matrix $T(\lambda)=A-\lambda B+\frac{\lambda}{\lambda-\sigma}C$ has a single pole $\sigma=\kappa/m$, i.e., $\Xi = \{\kappa/m\}$. In the results below we take $m=1$ and $\kappa=1$.
  \item The \textsf{gun} problem models a radio-frequency gun cavity \cite{Liao:2010:NRI}. Here, $T(\lambda)=K-\lambda M+i\sqrt{\lambda-\kappa^2_1}W_1+i\sqrt{\lambda-\kappa^2_2}W_2$, where we have used $\kappa_1=0$ and $\kappa_2=108.8774$. The nonlinearity comes from the square roots, and in this case $\Xi=(-\infty,\kappa_2^2)$. Figure~\ref{fig:gun} shows a pictorial representation of this problem.
  \item The \textsf{dimer} problem is taken from~\cite{Araujo:2020:CSR} and corresponds to the analysis of scattering resonances (TM polarization) of a dimer nano-structure (two disks coated with gold). The differential equation is expressed as $\Delta u+\varepsilon(x,\omega)\omega^2u=0$, where $\omega$ is the eigenvalue. A Drude-Lorentz model is used for the relative permittivity $\varepsilon(x,\omega)$, which approximates it as a rational function. Hence, the nonlinear eigenproblem to be solved is rational, with $T(\omega)=A_0-\omega^2A_1-\omega^2\varepsilon(x,\omega)A_2$. In the case of gold, the rational representation of $\varepsilon$ has 12 poles, which are depicted in Figure~{\ref{fig:dimer}}.
\end{itemize}
Both \textsf{delay} and \textsf{loaded\_string} have real eigenvalues, so the computation has been performed in real arithmetic. In the case of \textsf{gun} and \textsf{dimer} the problem matrices are complex, as well as the eigenvalues, and hence the solvers have been configured to use complex arithmetic.

\begin{table}%
\caption{Description of the test problems used in the computational experiments. The problem dimension is given, as well as the region where eigenvalues are sought (it is used only by the Interpol and NLEIGS solvers). In the tests, \emph{nev} eigenvalues are requested around the value $\sigma$ with a tolerance \emph{tol}.}
\label{tab:nepproblems}
\begin{minipage}{\columnwidth}
\begin{center}
\begin{tabular}{lcccccc}
\toprule
name&dim&region&\emph{nev}&$\sigma$&\emph{tol}\\
\midrule
\textsf{delay} 4M	&$2^{22}$& $[-100,50]$		&10& 0 & $1\times 10^{-8}$\\
\textsf{delay} 100K	&$100000$& $[-100,50]$		& 5& 1 & $1\times 10^{-6}$\\
\textsf{loaded\_string} 4M	&$2^{22}$& $[4,800]$ 		& 9& 10& $1\times 10^{-13}$\\
\textsf{loaded\_string} 200K	&$200000$& $[4,800]$ 		& 9& 10& $1\times 10^{-8}$\\
\textsf{gun}		&  9956  & see Fig.~\ref{fig:gun} 	&10&$65000+500\mathrm{i}$& $1\times 10^{-8}$ \\
\textsf{dimer}		& 173725&$[-1,20]\times[-2,0]$ & 20 &$5.3-0.25\mathrm{i}$ & $1\times 10^{-8}$\\
\bottomrule
\end{tabular}
\end{center}
\end{minipage}
\end{table}%


\begin{figure}
\centering
  \begin{tikzpicture}[xscale=.65]
    \tikzstyle{interp}=[fill=blue];
    \tikzstyle{singul}=[fill=orange];
    \tikzstyle{eigen}=[red,thick];
    \draw[thin] (1,0) -- (-3.5,0);
    \draw[violet,thick,fill=violet!10,opacity=.3] (1,-0.00008) -- (9.64,-0.00008) -- (9.64,2.4) -- (5.6,2.4) -- cycle ;
    \foreach \i/\j in {   
        1/0, 9.64/2.4, 1.54/0.28, 3.91/0, 1.05/0.027, 9.64/0.00041,
        1.15/0, 2.44/0.75, 1.01/0.0074, 5.60/2.4, 1.22/0.117, 1.95/0,
        1.03/0, 7.23/0, 3.71/1.42, 1.00/0, 1.41/0
    } \path[interp] (\i,\j) circle [radius=.06];
    \foreach \i in {   
        0.948, 0.254, 0.841,
        0.920, -0.902, 0.659, 0.939, -2.96, -0.155,
        0.889, 0.760, 0.945, 0.484
    } \path[singul] (\i,0) circle [radius=.06];
    \foreach \i/\j in {   
        4.36/0.037, 6.03/0.396, 6.18/0.0115, 6.48/0.0026, 3.90/0.0005,
        3.85/0.0034, 6.65/0.037, 3.54/0.00029, 3.51/0.0016, 6.95/0.0037,
        6.99/0.0029, 7.01/0.0026, 7.07/0.024, 7.86/0.015, 6.96/2.25,
        8.50/0.0069, 8.53/0.0022, 7.76/2.20, 1.79/0.000052, 8.79/0.011,
        8.79/0.0798, 8.56/2.043, 9.20/0.048, 9.45/0.344 }
      \draw[eigen] (\i-.075,\j-.075) -- +(.15,.15) (\i-.075,\j+.075) -- +(.15,-.15);
  \end{tikzpicture}
\caption{\label{fig:gun}Graphical representation of the \textsf{gun} problem. The shaded region is the target set $\Sigma$. The nodes (blue points) and poles  (orange points) used in the NLEIGS solver are Leja-Bagby pairs, picked from the discretization of the boundary $\delta\Sigma$ and the singularity set $\Xi$, respectively. Red crosses are the eigenvalues lying inside the region.}
\end{figure}
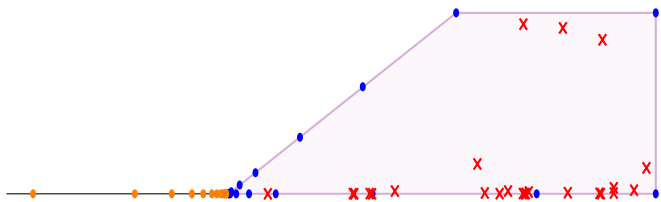
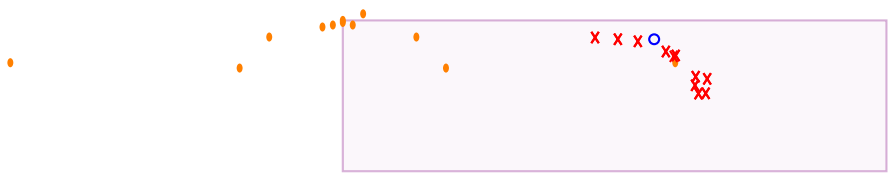
\begin{figure}
\centering
  \begin{tikzpicture}[xscale=.65]
    \tikzstyle{interp}=[fill=blue];
    \tikzstyle{singul}=[fill=orange];
    \tikzstyle{eigen}=[red,thick];
    \tikzstyle{target}=[blue,thick];
    \draw[violet,thick,fill=violet!10,opacity=.3] (-1,-2) -- (-1,0) -- (10,0) -- (10,-2) -- cycle ;
    \foreach \i/\j in {   
       -1.0/0, -0.588569/0.087417, 1.08757/-0.631936,
       -1.0/-0.0268586, -1.41143/-0.087417, -3.08757/-0.631936,
       -0.798753/-0.0610652, 0.48835/-0.220443, 5.72676/-0.560989,
       -1.20125/-0.0610652, -2.48835/-0.220443, -7.72676/-0.560989
    } \path[singul] (\i,\j) circle [radius=.06];
    \foreach \i/\j in {   
   5.536677/-0.412646,     5.536677/-0.412646,     4.970181/-0.277212,     4.970181/-0.277212,     5.693849/-0.476371,     5.736075/-0.465265,     5.736075/-0.465265,     4.565566/-0.249631,     4.565566/-0.249631,     6.139740/-0.746312,     6.139740/-0.746312,     6.126365/-0.860177,     6.126365/-0.860177,     6.196042/-0.970651,     6.196042/-0.970651,     6.372662/-0.774917,     6.372662/-0.774917,     4.104008/-0.226649,     4.104008/-0.226649,     6.343002/-0.966715,     6.343002/-0.966715        
    } 
      \draw[eigen] (\i-.075,\j-.075) -- +(.15,.15) (\i-.075,\j+.075) -- +(.15,-.15);
	      \foreach \i/\j in {   
5.3/-0.25      
    }  \draw[target] (\i,\j) ellipse (1mm and .65mm);
  \end{tikzpicture}
\caption{\label{fig:dimer}Graphical representation of the \textsf{dimer} problem. The orange points are the 12 poles of the rational function. The shaded part is the region used in the NLEIGS solver, while the blue circle is the target value used in NLEIGS as well as the other solvers. Red crosses are the 20 eigenvalues computed by NLEIGS (there are more eigenvalues inside the region, and other solvers may not return exactly the same ones).}
\end{figure}

All problems have been represented in the split form~\eqref{eq:split}. The accuracy of the computed eigenpairs is assessed with the scaled residual $\eta(x,\lambda)$, {\eqref{eq:sressplit}}, using $\infty$-norms for practical computation of matrix norms. Table \ref{tab:nepaccuracy} shows the maximum value of the scaled residual obtained by the different solvers on the problems of Table~\ref{tab:nepproblems}. A few comments are required to understand the results:
\begin{itemize}
\item The large problem size 4M for \textsf{delay} and \textsf{loaded\_string} has been used for the scalability tests (see below). In these cases, the computation uses an iterative linear solver, in particular, Bi-CGStab with AMG preconditioning (Hypre). The rest of executions use an LU factorization computed with MUMPS.
\item In the \textsf{gun} problem, the RII solver is run with deflation threshold activated, cf.\ \S\ref{sec:deflation}.
\item In RII, we use the Hermitian variant of the scalar equation~\eqref{eq:rii-scalar} (\texttt{NEPRIISetHermitian}) whenever the problem is Hermitian, i.e., in \textsf{loaded\_string} and \textsf{delay}.
\item The Newton-type solvers (SLP, RII, N-Arnoldi) compute \emph{nev} eigenvalues at most, while Krylov-type solvers (Interpol, NLEIGS) may return more than \emph{nev} eigenvalues in some cases. In particular, for the \textsf{delay} 100K problem Interpol and NLEIGS compute 6 eigenvalues, and for the \textsf{gun} problem NLEIGS computes 13.
\item In Krylov-type solvers (Interpol, NLEIGS), the \texttt{ncv} parameter (maximum dimension of the subspace, \texttt{NEPSetDimensions}) is set to the default $\max\{2\cdot nev,nev+15\}$, and the number of iterations listed in Table~{\ref{tab:nepaccuracy}} refers to the number of \emph{restarts}. For instance, in the \textsf{delay} problem a single restart cycle was enough, so the solver built 20 Krylov vectors (see the column `lin.~solves' in the table).
\end{itemize}

\begin{table}%
\caption{Results for the different solvers corresponding to executions using 16 MPI processes. Apart from the method name, we show the number of iterations, the number of required linear solves, the time required to build the preconditioner, the total time, and the scaled residual error.}
\label{tab:nepaccuracy}
\begin{minipage}{\columnwidth}
\begin{center}
\begin{tabular}{l|cccccccccc}
\toprule
test problem & method   & iter &lin.\ solves & PC build & time & $\eta(x,\lambda)$\\
\midrule              
\textsf{delay} 100K     & SLP		&  2	&  16	&  0.7	&  3.0	&$5\times 10^{-9}$\\
                        & RII		&  1	&  15	&  0.7	&  2.9	&$4\times 10^{-9}$\\
                        & N-Arnoldi	&  1	&  15	&  0.7	&  2.7	&$1\times 10^{-8}$\\
                        & Interpol  	&  1	&  20	&  0.7 	&  3.6	&$4\times 10^{-11}$\\
                        & NLEIGS    	&  1	&  20	&  0.7	&  3.6	&$6\times 10^{-14}$\\
\midrule
\textsf{loaded\_string} 200K& SLP	&  6	&  96	&  4.3	& 31.7	&$4\times 10^{-10}$\\
                        & RII		&  1	&  45	&  1.4	& 14.5	&$2\times 10^{-7}$\\
                        & N-Arnoldi	&  1	&  45	&  1.4	& 14.6	&$5\times 10^{-6}$\\
                        & Interpol      &  10	&  96	&  1.4	&  31	&$2\times 10^{-9}$\\
                        & NLEIGS   	&  3	&  41	&  1.4	& 13.4	&$2\times 10^{-10}$\\
\midrule
\textsf{gun}            & SLP		&  7	& 121	& 1.5	& 2.1	&$5\times 10^{-7}$\\
                        & RII		&  40	& 88	& 4.7	& 5.7	&$5\times 10^{-6}$\\
                        & N-Arnoldi	&  42	&  123	& 7.0	& 3.0	&$2\times 10^{-6}$\\
                        & NLEIGS    	&  3	& 39	& 0.4	& 0.6	&$1\times 10^{-11}$\\
\midrule
\textsf{dimer}          & SLP		&  18	& 401	& 21.6	& 55.5	&$1\times 10^{-9}$\\
                        & RII		&  89	& 249	& 49	& 329	&$1\times 10^{-8}$\\
                        & N-Arnoldi	&  115	& 571	& 90	& 249	&$9\times 10^{-9}$\\
                        & Interpol    	&  3	& 75	& 2.0	& 8.9	&$2\times 10^{-5}$\\
                        & NLEIGS    	&  2	& 58	& 2.0	& 5.4	&$2\times 10^{-14}$\\
\bottomrule
\end{tabular}
\end{center}
\end{minipage}
\end{table}%

From the table, we can see that Newton-based methods (SLP, RII, N-Arnoldi) are always slower than the polynomial or rational interpolation methods. In the case of the largest problem size (4M), they do not succeed to give the solution in a reasonable time. Among the classical solvers, we can see that N-Arnoldi is significantly faster than RII in the \textsf{gun} and \textsf{dimer} problems, which is to be expected from the accelerating effect of the expanding subspace in N-Arnoldi. Still, the SLP method beats both of them in these two problems.

As pointed out in \S\ref{sec:interpol}, the Interpol solver is restricted to real eigenvalues. In the case of \textsf{delay} we can hardly see any difference between Interpol and NLEIGS, as expected since NLEIGS with no singularities (poles set to infinity) behaves like a polynomial interpolation method. In contrast, \textsf{loaded\_string} already displays a significant gain of NLEIGS compared to Interpol.


\begin{table}%
\caption{Results for different variants of the NLEIGS solver with 16 MPI processes. The information shown is the same as in Table~\ref{tab:nepaccuracy}.}
\label{tab:nepnleigs}
\begin{minipage}{\columnwidth}
\begin{center}
\begin{tabular}{l|cccccccccc}
\toprule
test problem & NLEIGS variant & iter &lin.\ solves & PC build & time & $\eta(x,\lambda)$\\
\midrule              
\textsf{gun}   & TOAR basis &  3	& 39	& 0.4	& 0.6	&$1\times 10^{-11}$\\
               & full basis &  3	& 39	& 0.4	& 0.6	&$9\times 10^{-13}$\\
               & two-sided  &  4	& 102	& 0.4	& 1.1	&$6\times 10^{-13}$\\
\midrule
\textsf{dimer} & TOAR basis &  2	& 58	& 2.0	& 5.4	&$2\times 10^{-14}$\\
               & full basis &  2	& 58	& 2.0	& 7.2	&$2\times 10^{-14}$\\
               & two-sided  &  2	& 126	& 2.0	& 21	&$2\times 10^{-14}$\\
\bottomrule
\end{tabular}
\end{center}
\end{minipage}
\end{table}%

We have carried out a few additional runs to compare the different NLEIGS variants: TOAR basis (the default), full basis, and two-sided. Table~\ref{tab:nepnleigs} shows the results. In the \textsf{gun} problem, the overhead of the full basis version is negligible (even though the number of terms in the linearization is quite large, $d=15$) because the orthogonalization time is small. In contrast, for the \textsf{dimer} problem we see that there is a significant benefit in using the TOAR basis variant instead of working with the full basis. In this case both the number of terms ($d=12$) and the problem size are large. Regarding the two-sided variant, we observe that in addition to the overhead attributable to using full basis, there is an extra cost needed for the approximation of the left eigenspace. A related comment is that parallel operations involving matrix transposes are always less efficient than the non-transposed counterparts, due to the fact that matrices are stored by rows.

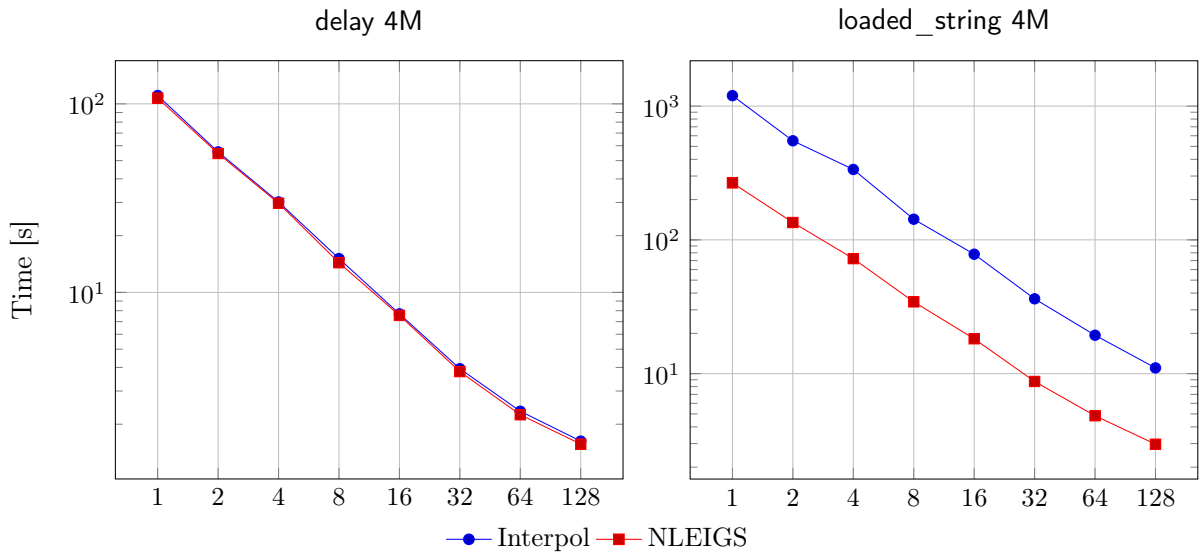
\begin{figure}
\centering
\resizebox{\textwidth}{!}{
\begin{tikzpicture}
\begin{loglogaxis}[
title={\textsf{delay 4M}},
ylabel={Time [s]},
grid=major,
log basis x=2,
xtick={1,2,4,8,16,32,64,128},
xticklabels={1,2,4,8,16,32,64,128},
ticklabel style={font=\small},
legend style={draw=none, legend columns=4, legend to name=legendname1,font=\small,cells={anchor=west}}
]
\addplot coordinates { 
(  1, 1.1072e+02)
(  2, 5.5617e+01)
(  4, 3.0185e+01)
(  8, 1.5120e+01)
( 16, 7.6964e+00)
( 32, 3.9385e+00)
( 64, 2.3433e+00)
(128, 1.6271e+00)
};
\addplot coordinates { 
(  1, 1.0734e+02)
(  2, 5.4539e+01)
(  4, 2.9728e+01)
(  8, 1.4369e+01)
( 16, 7.5544e+00)
( 32, 3.8028e+00)
( 64, 2.2445e+00)
(128, 1.5689e+00)
};
\legend{Interpol,NLEIGS}
\end{loglogaxis}
\end{tikzpicture}
\hfill
\begin{tikzpicture}
\begin{loglogaxis}[
title={\textsf{loaded\_string 4M}},
ylabel={},
grid=major,
log basis x=2,
xtick={1,2,4,8,16,32,64,128},
xticklabels={1,2,4,8,16,32,64,128},
ticklabel style={font=\small},
]
\addplot coordinates { 
(  1, 1.1973e+03)
(  2, 5.5027e+02)
(  4, 3.3603e+02)
(  8, 1.4267e+02)
( 16, 7.8134e+01)
( 32, 3.6250e+01)
( 64, 1.9373e+01)
(128, 1.1034e+01)
};
\addplot coordinates { 
(  1, 2.6628e+02)
(  2, 1.3478e+02)
(  4, 7.2401e+01)
(  8, 3.4402e+01)
( 16, 1.8221e+01)
( 32, 8.7387e+00)
( 64, 4.8433e+00)
(128, 2.9687e+00)
};
\end{loglogaxis}
\end{tikzpicture}
}
\ref*{legendname1}
\caption{\label{fig:nepscal}Execution time (in seconds) with up to 128 MPI processes for the problems \textsf{delay} (left) and \textsf{loaded\_string} (right), solved with Interpol and NLEIGS. The execution parameters are shown in Table~\ref{tab:nepproblems}.}
\end{figure}

We conclude this section by analyzing the parallel scalability of the two interpolation solvers. Figure~\ref{fig:nepscal} shows the execution time of both Interpol and NLEIGS when solving the large problem sizes (4M) with increasing number of MPI processes. The plots show a very good scaling up to 128 processes.

\section{Conclusions\label{sec:concl}}

This paper collects in one place all the relevant details related to the solvers for the nonlinear eigenvalue problem that we have been developing in SLEPc in the last years. These solvers constitute the \texttt{NEP} module of SLEPc, which provides a flexible user interface to specify the nonlinear problem in different application contexts. We have discussed the details of SLP, RII, N-Arnoldi, Interpol and NLEIGS, including important issues such as deflation or parallelization. The available user options have been mentioned as well, as a way of brief users manual. We have also illustrated the performance of the implemented methods when solving a set of representative problems. The results show that good accuracy can be attained with all the methods, with more or less efficiency, and that very large scale problems can be addressed thanks to the good parallel scalability.

Apart from a mere description of the implementations, this paper includes two new developments: (1) a version of NLEIGS that stores the Krylov basis in compact format (\`a la TOAR), in the same spirit as previous works such as~\cite{Campos:2016:PKS} for polynomial eigenproblems and~\cite{Dopico:2018:CRK} for rational eigenproblems; and (2) a two-sided version of NLEIGS that can be used to compute left eigenvectors, which are required, e.g., to approximate the resolvent $T^{-1}(z)$. Our two-sided NLEIGS method operates without explicitly building the linearization matrices, but in contrast it stores the Krylov basis explicitly, as opposed to the variant of~\cite{Lietaert:2018:CTK}. We have also derived the details for a practical implementation of Effenberger deflation in the context of classical iterations (SLP, RII, and N-Arnoldi).

Research on nonlinear eigenvalue problems is still very active, and we foresee that novel methods will be proposed in the next years. Our intention is to continue adding solvers to the \texttt{NEP} module, so that users have a wide range of choices from which to select the most appropriate one for each particular application. As an example, we recently finished the implementation of the two-sided variant of SLP with non-equivalence deflation, that was included in version 3.13 of SLEPc.
Examples of recently proposed methods that could be incorporated in the future are~\cite{Gavin:2018:FEN}, \cite{Mele:2018:RTI}, \cite{Xue:2018:BPH}.
On the other hand, the area of nonlinear eigenvector problems (that is, problems that depend nonlinearly on the eigenvector) remains as a topic of future investigation.

During the review process, the preprint \cite{Guttel:2020:RRA} appeared, which proposes several improvements over the rational interpolation solvers of \S\ref{sec:nleigs}, such as repeating the poles cyclically in Leja-Bagby order instead of adding extra poles at infinity. We will adapt these ideas to our \texttt{NEP} solvers in forthcoming versions of SLEPc.

\section*{Acknowledgements}
The authors would like to thank the anonymous referees whose constructive comments helped improve the presentation.
The computational experiments of section \ref{sec:performance} were carried out on the supercomputer Tirant III belonging to Universitat de Val\`encia.


\end{document}